\PassOptionsToPackage{table,xcdraw}{xcolor}

\documentclass[acmsmall,screen]{acmart}

\AtBeginDocument{%
  \providecommand\BibTeX{{%
    \normalfont B\kern-0.5em{\scshape i\kern-0.25em b}\kern-0.8em\TeX}}}

\setcopyright{none}
\renewcommand\footnotetextcopyrightpermission[1]{}
\usepackage{xspace}
\usepackage[inline]{enumitem}
\usepackage{adjustbox}
\usepackage{multirow}

\newcommand{\eg}{\textit{e.g.,}\xspace}
\newcommand{\ie}{\textit{i.e.,}\xspace}

\newcommand\rqZero{RQ1: How do developers perceive and experience conflicts during code review?\xspace}
\newcommand\rqOne{RQ2: What are conflicts during code review like?\xspace}
\newcommand\rqTwo{RQ3: What are the positive and negative consequences of conflicts during code review?\xspace}
\newcommand\rqThree{RQ4: What factors intervene in the conflict dynamics?\xspace}
\newcommand\rqFour{RQ5: What strategies do developers use to prevent and manage conflicts?\xspace}
\newcommand{\qu}[1]{``#1''\xspace}
\newcommand{\qud}[2]{\qu{#1}~[D#2]\xspace}

\begin{document}

\title{Interpersonal Conflicts During Code Review}
\subtitle{Developers' Experience and Practices}

\author{Pavl\'{i}na Wurzel Gon\c{c}alves}
\email{p.goncalves@ifi.uzh.ch}
\orcid{0000-0002-2231-054X}
\affiliation{%
  \institution{University of Zurich}
  \streetaddress{Binzm\''{u}hlestrasse 14}
  \city{Zurich}
  \state{ZH}
  \country{Switzerland}
  \postcode{8050}
}

\author{G\"{u}l \c{C}al{\i}kl{\i}}
\email{HandanGul.Calikli@glasgow.ac.uk}
\orcid{0000-0003-4578-1747}
\affiliation{%
  \institution{University of Glasgow}
  \city{Glasgow}
  \state{Scotland}
  \country{United Kingdom}
}

\author{Alberto Bacchelli}
\email{bacchelli@ifi.uzh.ch}
\orcid{0000-0003-0193-6823}
\affiliation{%
  \institution{University of Zurich}
  \streetaddress{Binzm\''{u}hlestrasse 14}
  \city{Zurich}
  \state{ZH}
  \country{Switzerland}
  \postcode{8050}
}

\begin{abstract}
Code review consists of manual inspection, discussion, and judgment of source code by developers other than the code's author.
Due to discussions around competing ideas and group decision-making processes, interpersonal conflicts during code reviews are expected.
This study systematically investigates how developers perceive code review conflicts and addresses interpersonal conflicts during code reviews as a theoretical construct. Through the thematic analysis of interviews conducted with 22 developers, we confirm that conflicts during code reviews are commonplace, anticipated and seen as normal by developers. Even though conflicts do happen and carry a negative impact for the review, conflicts---if resolved constructively---can also create value and bring improvement.
Moreover, the analysis provided insights on how strongly conflicts during code review and its context (\ie code, developer, team, organization) are intertwined. Finally, there are aspects specific to code review conflicts that call for the research and application of customized conflict resolution and management techniques, some of which are discussed in this paper.\\
Data and material: \url{https://doi.org/10.5281/zenodo.5848794}
\end{abstract}



\maketitle

\section{Introduction}
\label{sec:intro}
Code review, is a manual inspection of source code by developers other than the code's author and is considered a valuable tool for finding defects and improving software quality~\cite{Ackerman:1989}.
Unlike the highly structured code inspection formalized almost fifty years ago~\cite{Fagan:1976}, the most widespread form of code review nowadays (also known as modern code review~\cite{bacchelli2013expectations}) is a change-based event, requiring short-term reactions from reviewers and ongoing feedback from colleagues~\cite{sadowski2018modern}. Through discussion, a decision is reached on whether a code change can be merged into the codebase or whether it should be reworked or rejected~\cite{davila2021systematic}.

Effective code reviewing requires understanding each others' code and comments and making one's own understandable~\cite{bacchelli2013expectations}--challenging activities that pose cognitive load on the developer and reviewer(s)~\cite{Thongtanunam:2015, Baum:2016a, Baum:2017}. Moreover, modern code review benefits from discussing competing ideas~\cite{bacchelli2013expectations}, therefore in addition to misinterpretations about code and comments, a misalignment in developers' goals, priorities, ideas, and perspectives may exist. These features of code review make \textit{interpersonal conflicts} likely to happen~\cite{hartwick2002conceptualizing}.

\citet{hartwick2002conceptualizing} define an interpersonal conflict as ``a dynamic process that occurs between interdependent parties as they experience negative emotional reactions to perceived disagreements and interference with the attainment of their goals.''

Interpersonal conflicts have been studied within the context of open-source software (OSS)~\cite{filippova2016effects, filippova2015mudslinging, huang2016effectiveness} and Information Systems (IS) development~\cite{barki2001interpersonal}. In the latter context, \citet{barki2001interpersonal} have described each conflict situation according to three components: \textit{negative emotions}, cognitive \textit{disagreement}, and \textit{interference of behavior}.

\begin{figure}
  \fbox{\includegraphics[width=0.8\textwidth] {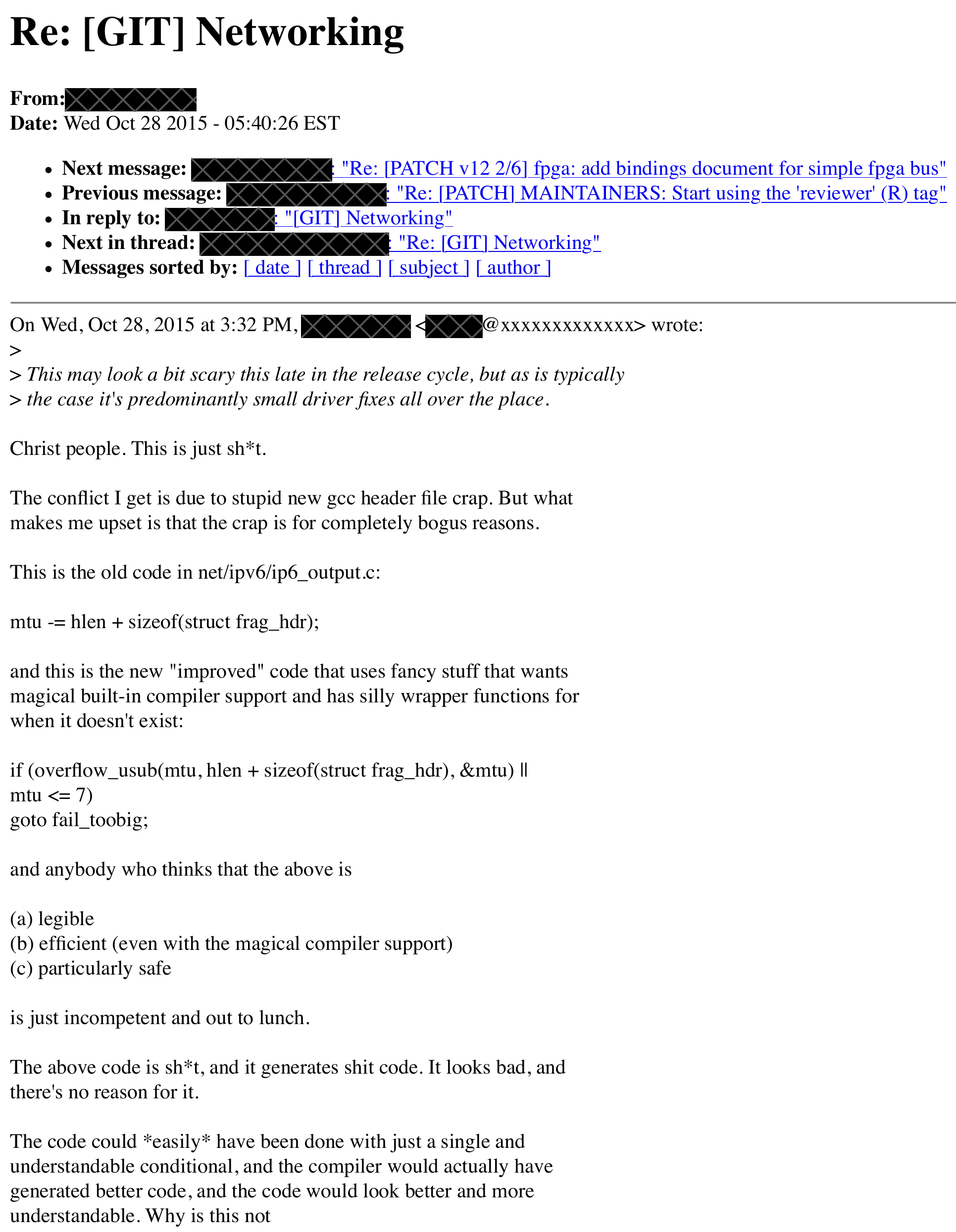}}
  \caption{Comment made in the Linux mailing list about a pull request for Linux version 4.3~\cite{LinusTorvaldsEmail}.}
\label{fig:fig1}
\end{figure}

An example of a potential interpersonal conflict during code review can be found in the Linux Kernel Mailing List (LKML)~\cite{LKML}. Figure \ref{fig:fig1} shows an email where the sender comments on a pull request in LKML~\cite{LinusTorvaldsEmail}.
This email message contains evidence of:

\begin{enumerate}
 \item \textit{negative emotions} (\eg ``\ldots what makes me \textit{upset} is that the crap is for completely bogus reasons''),
 \item \textit{disagreements} (\eg ``\ldots the code could easily have been done with just a single and understandable conditional''), and
 \item \textit{interference} (\ie writing an email to a mailing list with a harsh language likely to let the pull request's author to experience negative emotions).
\end{enumerate}

However, the use of harsh language is not the only form of how conflict can take place. For instance, empirical evidence shows that developers can get frustrated if reviewers block their code change requests by withholding the code change's acceptance too long or by asking for excessive modifications~\cite{egelman2020pushback}. In general, software practitioners repeatedly refer to the adverse effects of conflicts during code review~\cite{insufferable, ruining, linus}. For example, OSS developers can stop contributing to a project because of conflicts during the review process~\cite{huang2016effectiveness}. Yet, the literature supports that conflicts can be constructive besides being destructive by bringing change and solutions to problematic situations~\cite{deutsch1994constructive}.

\begin{figure}[t]
      \centering
      \includegraphics[width=0.85\columnwidth]{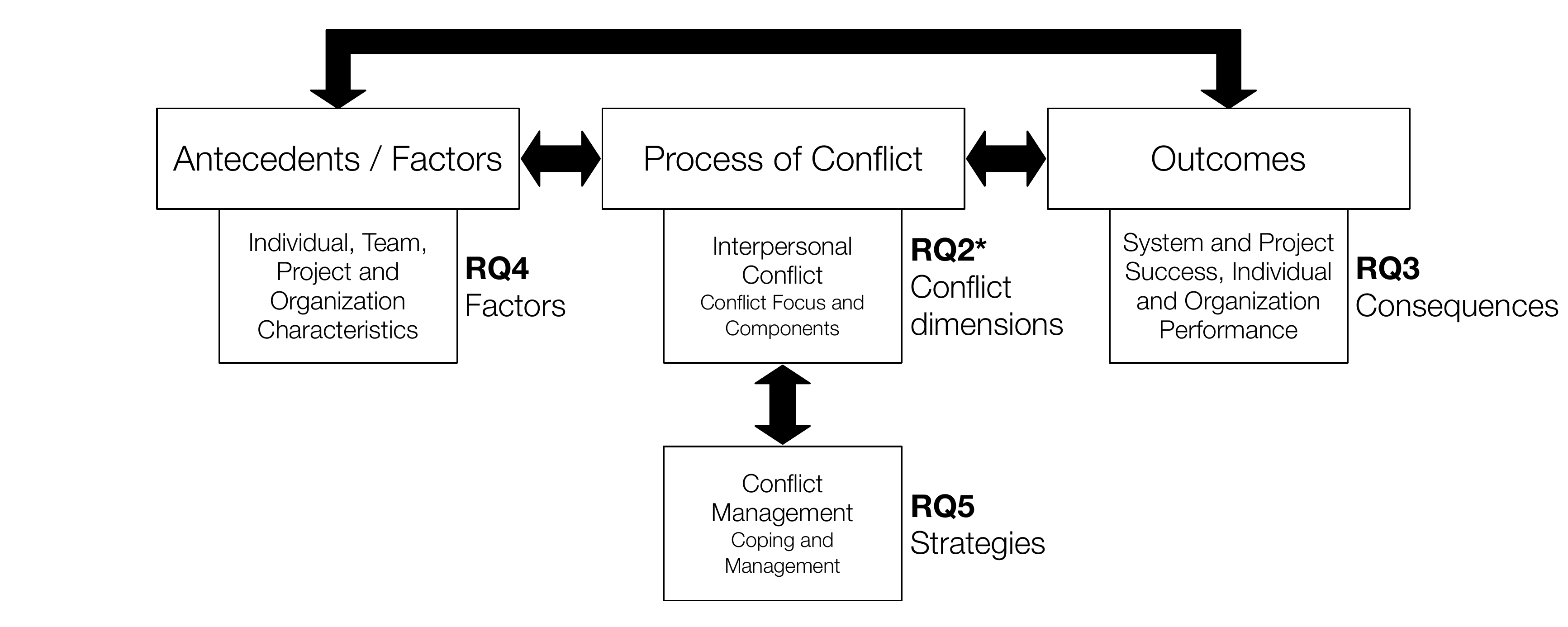}
      \caption{A Conceptual Framework to Investigate Code Review Conflicts (adapted from Barki and Hartwick~\citet{barki2001interpersonal}).}
       \begin{flushleft}
        \textsuperscript{* The conceptual framework employs the two dimensional construct by  \citet{hartwick2002conceptualizing} to define \emph{Interpersonal Conflict}.}
       \end{flushleft}
      \label{fig:conceptual_framework}
\end{figure}

Analyzing the aforementioned conflict components (\ie \textit{negative emotions}, \textit{interference}, and \textit{disagreement}) can help detect when a conflict is happening and is a first step towards measuring the risk of conflicts. Measurement of the conflict components can lead to a proper assessment of the individuals' perceived level of interpersonal conflicts ~\cite{barki2001interpersonal}, which facilitates both the assessment of the severity of conflicts and the formulation of appropriate resolution and management strategies. The assessment of the level at which conflicts are perceived is also crucial to evaluate the effectiveness of the conflict resolution and management strategies when put into practice. Moreover, knowing the \textit{conflict focus} (\eg whether the conflict is about the work to be done or how the work needs to be done) is essential to formulate conflict resolution and management strategies specific to the situation. Hartwick and Barki's construct of interpersonal conflicts consists of the two dimensions (\textit{conflict components} and \textit{conflict focus}) and provides a basis for in-depth analysis of interpersonal conflicts~\cite{hartwick2002conceptualizing}. Moreover, the development of conflict resolution and management techniques requires understanding the interaction of the conflict process with the conflict's antecedents and outcomes, which is depicted in the conceptual framework (Figure \ref{fig:conceptual_framework}) proposed by \citet{barki2001interpersonal}.

Despite the prevalence of code review in practice and its nature that can spark conflicts, few studies touch upon this topic and only focus on a \textit{single aspect of conflicts} such as pushback~\cite{egelman2020pushback} and profanity~\cite{squire2015floss, schneider2016differentiating}. Conflict resolution with constructive suggestions was empirically identified as an effective strategy to prevent developers from leaving OSS projects~\cite{huang2016effectiveness}. However, studies are needed to further understand how conflicts during code reviews happen and developers' motivations and behavior in these situations.

In this paper, we present an in-depth investigation of interpersonal conflicts in code review as a theoretical construct and we systematically describe: (1) components and focus of conflicts, (2) antecedents of conflicts, (3) consequences of conflicts, and (4) developers' strategies for conflict resolution and management. To conduct our investigation, we carried out 22 interviews and analyzed our findings referring to the conceptual framework we adapted from \citet{hartwick2002conceptualizing}.

Our findings indicate that \textbf{conflicts are common} during code reviews and \textbf{perceived as natural}, mainly due to the exchange of competing ideas. Moreover, using conflicts as a learning and improvement opportunity aids the fulfillment of code review goals; therefore, conflicts are \textbf{not to be completely prevented but managed}. While developing strategies for constructive resolution of conflicts, one should consider that \textbf{code review is strongly intertwined with its context}, which can comprise \textit{code}, \textit{developer}, \textit{team}, and \textit{organization}. Moreover, \textbf{conflicts in code review require strategies that are specific to its specific context}. Thus, new tools and techniques need to be devised rather than using existing off-the-shelf techniques.

\section{Background and Related Work}
\label{sec:related_work_general}
This section provides general background on interpersonal conflicts in organizations, Information Systems (IS), software engineering, and conflict management. Finally, we provide information on code review practices and the type of code review we consider in this study and we mention studies on social aspects of code review related to conflicts and studies addressing specific aspects of code review conflicts (\eg pushback~\cite{egelman2020pushback}, profanity~\cite{schneider2016differentiating, squire2015floss}).

Interpersonal conflict has been studied in diverse fields, including psychology, management, and communication.
Despite being an unpleasant experience often with adverse effects, conflicts also present opportunities for improvement and innovation~\cite{coleman2012constructive}.

Studies on apparent conflicts are prevalent in the literature~\cite{kolb1992hidden}. However, conflicts can also be \textit{hidden}, especially in organizations, due to the need to keep a professional appearance. \citet{Dubinskas:1992} defines \textit{hidden conflicts} as conflicts that often are invisible to the conflict's own members, which in turn hides such conflicts from the analysis. Analyzing hidden conflicts is crucial to understand the origins and consequences of organizational conflict. Hidden conflicts can threaten or preserve the order in organizations as well as leading to incremental changes in the organizations through the adoption of new routines. Moreover, hidden conflict resolution requires strategies that might be informal and different from traditional ways of conflict intervention~\cite{Kolb:1992, Bartunek:1992}.

\subsection{Conflicts in Information Systems and Software Engineering}
\label{conflicts_IS_SWE}
There is evidence on the existence of conflicts between IS users and developers~\cite{barki2001interpersonal}, and within distributed development teams~\cite{alqhtani2014proposal}, including OSS development teams~\cite{filippova2016effects, filippova2015mudslinging} and collaborative environments~\cite{torok2013opinions}, such as Wikipedia~\cite{yasseri2012dynamics, zhang2018conversations, jacquemin2008managing, sumi2011edit, kittur2007he}.
Conflicts can impair collaborators' productivity and the quality of their work~\cite{arazy2013stay, arazy2011information} and even lead them to leave the project~\cite{huang2016effectiveness}. Moreover, the presence of conflicts is related to lower process satisfaction, decreased system quality, and worse adherence to budget, schedule, and requirements~\cite{barki2001interpersonal}, as well as to a lower perception of project's performance and reduced developers' identification with their OSS team~\cite{filippova2016effects, filippova2015mudslinging}.

\citet{filippova2016effects,filippova2015mudslinging} categorized conflicts during OSS development as the task, process, norm, and affective conflicts. Differently from \citet{filippova2016effects,filippova2015mudslinging}, \citet{hartwick2002conceptualizing} defined an interpersonal conflict as a two-dimensional construct consisting of \textit{conflict focus} and \textit{components}. They also conducted a survey with IS staff and users to analyze correlations among the \textit{conflict components} (\ie \textit{negative emotions}, \textit{interference}, and \textit{disagreement}). Their findings serve as the empirical verification of the hypothesis stating that interpersonal conflict comprises all three components.  \citet{barki2001interpersonal} also conceptualized a framework that we adapt to conduct an in-depth analysis of interpersonal conflicts within the context of code review (see Figure~\ref{fig:conceptual_framework}). The conceptual framework depicts the interaction between the \emph{antecedents} and \emph{outcomes} of conflict as well as the \emph{process of conflict}.  The \emph{process of conflict} consists of \emph{interpersonal conflict}, defined by its two dimensions as well as by \emph{conflict management}.

\subsection{Conflict Management}
\label{sec:management}
An extensive body of knowledge covers principles and procedures for constructive conflict resolution from the perspective of management~\cite{coleman2012constructive, deutsch1994constructive} and individual coping strategies~\cite{ben2009coping}.
Some conflict resolution and management strategies proposed in the management literature are:  distributive bargaining, integrative negotiation, joint training, complex-systems change, and de-escalation~\cite{coleman2012constructive, deutsch1994constructive, ben2009coping}.

Several strategies to conflict handling and management have been investigated in the context of software development, such as compromising, dominating, avoiding, problem-solving, obliging, integrating, smoothing, rational explanation, social encouragement, forcing, confronting, or accommodating~\cite{gobeli1998managing, domino2003conflict, barki2001interpersonal, huang2016effectiveness, ben2009coping}. These studies underline that problem-focused, collaborative, and pro-active solutions relate to better conflict outcomes. While such general approaches and strategies have a broad applicability, specific contexts (\eg OSS development and collaborative work) may require specific strategies for conflict resolution~\cite{huang2016effectiveness}.

Formulating the collaborative problem-solving strategies for conflict management in Modern Code Review can have implications for the software engineering practice. Currently, studies in software development derive the conflict management strategies they investigate from the existing literature~\cite{gobeli1998managing, domino2003conflict, barki2001interpersonal} or manual inspection of software development artifacts, such as pull-request comments~\cite{huang2016effectiveness}.  Strategies that \citet{huang2016effectiveness} derived from pull-request comments can capture only communicational strategies that are directly observable in the review discussion text. However, such strategies cannot capture developers' interpretations and intents or strategies developers might use outside of the review discussions, such as in-person meetings.  To address these shortcomings, in this study, we conduct interviews with developers to investigate the strategies they use to prevent and manage conflicts during code review. This approach allows us to capture the subjective motivations and developers' approaches and we expect it to let us discover a broader range of conflict management strategies.

\subsection{Modern Code Review}
\label{sec:mcr}
The most widespread contemporary practice of code review (also known as Modern Code Review (MCR) ~\cite{bacchelli2013expectations, cohen2010modern}) consists in an informal and lightweight process that focuses on inspecting new proposed code changes rather than the whole codebase~\cite{rigby2013convergent}. 
During MCR, a developer (author) submits a code change that other developers (reviewers) inspect. The main goal of this process is to find as many issues as possible in the submitted code change and provide feedback the author needs to address before the change is accepted and put into production~\cite{baum2016factors}. The reviewers and the code change's author engage in an asynchronous discussion: The reviewers can ask for clarifications or recommend improvements to the author, who can reply to the comments and propose improvements.
This mechanism can contain the upload of new versions of the code change (\ie revised patches or iterations), which leads to an iterative process that finishes when all the reviewers are satisfied with the change or decide not to include it in production.

Besides being informal, lightweight, and asynchronous, MCR is tool-based. Among widely used code review tools are Gerrit~\cite{gerrit}, Microsoft CodeFlow~\cite{ms_codeflow}, Facebook's Phabricator~\cite{fb-phabricator}, and Atlassian Crucible~\cite{atlassian-crucible}. The core component of such code review tools is that the tool facilitates and reports the discussion between the reviewers and the author on the submitted code change~\cite{pascarella2018information}. GitHub pull-requests~\cite{github-pr} is another medium for MCR, especially popular with OSS projects. In the pull-based software development model, developers who contribute to the project (contributors) can clone the repository to make changes independently. When a set of changes is ready to be submitted to the main repository, a member of the project's core team (integrator) inspects the changes. If the changes are not satisfactory, the integrator requests more changes the contributor is supposed to implement. Once the integrator assesses the changes to be satisfactory, the contributors' changes are pulled to the main branch. Some OSS projects use mailing lists for code reviews. For instance, Linux Kernel Development heavily relies on mailing lists. A developer should send the code change (patch) to the appropriate subsystem maintainer through at least one of the Linux Kernel-related mailing lists~\cite{LKdev-ml:a, LKdev-ml:b}. The patch then usually gets comments from reviewers on how it can be improved. The developer addresses the problems the reviewer points out and informs the reviewer of the planned changes.

Besides code review tools, pull requests, and mailing lists, face-to-face discussions act as supplementary communication channels during code reviews. For instance, \citet{macleod2017code} report that the author and reviewer also engage in face-to-face discussions during code review at Microsoft, while most communication between author and reviewer occurs through the code review tool. Face-to-face discussions can occur in software development environments where teams work on an internal project in a company (\eg Microsoft developers), including teams working on an OSS project (\eg Mozilla developers). On the other hand, the code review process relies only on tools or mailing lists in an organically formed OSS community (\eg contributors to the Linux Kernel project).

Due to the variety in the code review process in different settings and the explorative nature of this study, we do not limit what type of policies developers follow or what communication and tools they use to perform code reviews.

\subsection{Social Code Review and Conflicts}
\label{related_work_CR}
Code review comprises social interactions about a technical problem. A number of studies in the software engineering literature focus on non-technical factors involved in the code review process. \citet{bosu2016process} showed that the quality of the code under review affects how the reviewers perceive the reliability, expertise, and trustworthiness of the contributor. Moreover, \citeauthor{bosu2016process} found that the code review process is a critical practice for creating successful projects' social underpinning because it helps developers form impressions about their teammates, which impact future collaborations besides future code reviews' outcomes.
\citet{baysal2013influence} found that human factors (\eg developers' experience and reviewer's load) significantly impact code review outcome. \citet{kononenko2015investigating} showed that the number of people involved in a code review is associated with the quality of the review itself.
Developers' experience, which was found to be associated with the code review outcome~\cite{baysal2013influence,kononenko2015investigating}, has been also used in code reviewer recommendation techniques~\cite{thongtanunam2015should, rahman2016correct}. \citet{thongtanunam2015should} proposed a code reviewer recommendation technique based on the developers' experience in reviewing files that are stored in close proximity with the code change to be reviewed; while the approach by \citet{rahman2016correct} is based on developers' experience in certain specialized technologies and external libraries. \citet{bosu2014impact} addressed developers' reputation in OSS environments, which is related to developers' experience. Authors found that OSS developers' reputation helps them get faster feedback and make their contributions more likely to be accepted.

Currently, few studies address some aspects of interpersonal conflicts during code review. \citet{egelman2020pushback} focus on pushback, defined as ``the perception of unnecessary interpersonal conflict in code review while a reviewer is blocking a change request.''  Other studies focus on specific aspects related to conflicts such as unfair treatment~\cite{german2018my}, and profanity and insults~\cite{schneider2016differentiating, squire2015floss} during code review. Empirical findings by \citet{german2018my} indicate that the majority of contributors to OpenStack, a large industrial OSS ecosystem, have experience with unfair treatment during code review. The study by \citet{huang2016effectiveness} analyses the effectiveness of three conflict management strategies on 170 GitHub projects' pull-based code review process, namely ``rational explanation'', ``constructive suggestion'', and ``social encouragement''. The authors identify ``constructive suggestions'' as an effective conflict management strategy after analyzing the review comments and outcome of a survey conducted with the GitHub projects' developers.

\section{Methodology}
This section presents our research questions and provides information on the employed analysis, ethics, and data storing besides the information on the interviews and characteristics of the 22 interviewed developers. Additional information and details on the methodology and other appendices are available in the supplementary material~\cite{material}.

\subsection{Research Questions}
\label{sec:RQs}
Both the study by \citet{egelman2020pushback} and the experience of practitioners as reported in publicly available resources~\cite{insufferable, ruining, linus} suggest that code review is a potential venue for interpersonal conflicts, especially because it requires frequent discussion of competing ideas.
Therefore, we start by asking:

\begin{description}[leftmargin=0.3cm]

  \item[\textbf{\rqZero}]
  RQ1 aims to capture how developers perceive conflicts related to code review (\eg whether conflicts during code review are an issue or whether they are specific in the context of code review).

\end{description}

To analyze conflicts during code review and formulate the remaining research questions, we rely on a conceptual framework of interpersonal conflicts (see Figure~\ref{fig:conceptual_framework}) adapted from \citet{barki2001interpersonal}. Human, organizational and technical factors act as antecedents of conflicts and are affected by the conflict outcomes, creating the ground on which further conflicts are (not) happening.

\begin{description}[leftmargin=0.3cm]

  \item[\textbf{\rqOne}]
  To answer RQ2, we refer to Hartwick and Barki's two-dimensional construct of interpersonal conflict~\cite{hartwick2002conceptualizing}. For this purpose, we analyze the data to investigate the focus of conflicts and the conflict components.

  \item[\textbf{\rqTwo}]
  Conflicts can also have the potential to be an opportunity for improvement~\cite{deutsch1994constructive}. Therefore, we are interested in whether developers see and experience any negative but also positive consequences of interpersonal conflicts in code review.

  \item[\textbf{\rqThree}]
  RQ4 focuses on understanding which characteristics of code review and its context contribute to conflicts, in terms of emergence, severity, and constructiveness. The conceptual framework by \citet{hartwick2002conceptualizing} formulates these characteristics as conflicts' \emph{antecedents}. However, the data we gathered from the interviews suggests that these characteristics can play many intervening roles and can also be moderators, mediators, and phenomena co-occurring with the conflicts that are related to how the conflict unfolds. Therefore, we investigate what \textit{factors} make conflicts more likely to happen or be beneficial rather than destructive.

  \item[\textbf{\rqFour}]
  Problem-focused and collaborative solutions are a way towards positive conflict outcomes~\cite{ben2009coping, barki2001interpersonal}. We aim to provide a developer-driven perspective on such problem-solving strategies, specifically in the context of code review. RQ5 investigates the strategies developers adopt and put into practice to prevent and manage conflicts.

\end{description}

\subsection{Interviews}
To answer our RQs, we conducted an exploratory study by analyzing data we obtained from semi-structured interviews with developers.  Appendix \ref{sec:appendixI} presents the interview structure.
Due to the complex nature of conflicts, to gather data for answering our research questions, we estimated each interview to last at least 45 minutes. We continued the interviews until we reached \emph{saturation}.

As this study explores a complex process, we aimed for \emph{code saturation} rather than \emph{meaning saturation}~\cite{hennink2017code}. In other words, we aimed to identify issues and topics related to conflicts to describe the landscape rather than to exhaustively describe and understand all possible insights within the codes and themes.  For the code saturation, around the 20th interview, the topics and issues became repetitive (even if details of the narratives could be extended), therefore we interrupted the interviews after 22 complete ones.\footnote{This number is aligned with the estimates for grounded theory studies, which state that between 20 to 30 interviews should be expected to conduct a saturated study~\cite{creswell2002educational}.}

Furthermore, the saturation is closely related to the sampling methods. Therefore, we have defined a 2-step sampling strategy (Section \ref{sec:sample}) to firstly identify what characteristics of developers provide a broader variety in topics that appear, secondly to hire these diverse profiles~\cite{onwuegbuzie2007call}.
As the interviews were progressing, it became noticeable that the narratives differed significantly, especially in three types of profiles--the different length of experience, different maturity of the code review process, and different environments of conducting code reviews. Therefore, in the second stage of sampling, we have tried to maximize the variety of profiles, covering more or less experienced developers, varying environments (company, Open Source Software community, academic), and more or less standardized modern code review process. We have used these characteristics to compose and describe the study sample (Section \ref{sec:sample}).
Finally, once the interviews were done, during the data analysis (described in Section \ref{sec:met:analysis}), we did not identify any undiscovered areas that seemed worthy investigating further.

\subsection{Ethics and Data Handling}
Before conducting the interviews, each participant signed a consent form (available in the online material).
We anonymized the interview transcripts before storing them on our institution's server. All non-anonymized data (\eg interview recordings) was deleted, and the external provider who transcribed the interviews signed a Non-Disclosure Agreement. We also ensured no direct connection between the anonymized transcripts and the consent forms the participants signed. Furthermore, our institution's Human Subjects Committee approved the methodology of our study.

\subsection{Sample}
\label{sec:sample}
We used a multi-stage purposeful sampling method~\cite{onwuegbuzie2007call}. In the first stage, we used convenience sampling, gathering data from developers who were available from our social and professional network to participate in the study. In the second stage, we used mixed purposeful sampling consisting of snowball and maximum variation sampling to ensure at the same time maximum possible variance in developer profiles with respect to `Code Review (CR) experience', `software development environment', and `CR process maturity'.
The sample is described in Table~\ref{tab:descriptives} with respect to these characteristics and frequency of experiencing conflicts.
Table~\ref{tab:casecount} describes the meaning of the categorical values in Table~\ref{tab:descriptives} and reports the number of developers in each category (\ie `count'). For example, in terms of `CR experience' we have three levels: \textit{entry} (novice developers or students), \textit{mid-career} (developers with responsibilities in the software development process), and \textit{supervisory} (for interviewees on a management level or who supervise and mentor others).

Concerning `CR process maturity' (Table~\ref{tab:casecount}), most interviewees (N=14) take part in processes with \textit{high} maturity (\ie systematic, structured, and integrated into the software development life cycle). Moreover, participants experience conflicts during code reviews at \textit{low} (N=2), \textit{occasional} (N=9), and \textit{regular} (N=11) frequencies.

\begin{table}
\caption{Descriptives of Interviewed Developers}

\label{tab:descriptives}
\begin{adjustbox}{width=\textwidth}
\begin{tabular}{lllll}

\textbf{ID} & \textbf{CR Experience} & \textbf{Software Development Environment}                      & \textbf{CR Process Maturity} & \textbf{Conflict Frequency} \\ \hline
D1          & Mid-Career                & Internal                                  & High                         & Regular                     \\
\rowcolor[HTML]{EFEFEF}D2          & Entry Level               & Internal                                  & Ad hoc                       & Occasional                  \\
D3          & Entry Level               & Internal                                  & High                         & Low                         \\
\rowcolor[HTML]{EFEFEF}D4          & Mid-Career                & Community, Academic (Researcher)          & Limited                      & Occasional                  \\
D5          & Mid-Career                & Internal                                  & Ad hoc                       & Regular                     \\
\rowcolor[HTML]{EFEFEF}D6          & Mid-Career                & Internal                                  & High                         & Occasional                  \\
D7          & Mid-Career                & Internal                                  & High                         & Occasional                  \\
\rowcolor[HTML]{EFEFEF}D8          & Supervisory               & Internal, Community                       & High                         & Regular                     \\
D9          & Supervisory               & Internal                                  & High                         & Regular                     \\
\rowcolor[HTML]{EFEFEF}D10         & Mid-Career                & Internal                                  & Limited                      & Occasional                  \\
D11         & Mid-Career                & Internal, Community                       & High                         & Low                         \\
\rowcolor[HTML]{EFEFEF}D12         & Entry Level               & Academic (Student)                        & Limited                      & Regular                     \\
D13         & Mid-Career                & Community, Academic (Researcher)          & Limited                      & Occasional                  \\
\rowcolor[HTML]{EFEFEF}D14         & Mid-Career                & Internal, Academic (Researcher)           & High                         & Regular                     \\
D15         & Supervisory               & Internal                                  & High                         & Regular                     \\
\rowcolor[HTML]{EFEFEF}D16         & Mid-Career                & Internal, Community                       & High                         & Occasional                  \\
D17         & Mid-Career                & Internal, Community                       & High                         & Regular                     \\
\rowcolor[HTML]{EFEFEF}D18         & Mid-Career                & Internal                                  & High                         & Regular                     \\
D19         & Supervisory               & Internal, Academic (Researcher)           & High                         & Regular                     \\
\rowcolor[HTML]{EFEFEF}D20         & Mid-Career                & Community, Academic (Researcher, Student) & Limited                      & Occasional                  \\
D21         & Supervisory               & Internal, Community                       & High                         & Regular                     \\
\rowcolor[HTML]{EFEFEF}D22         & Mid-Career                & Internal, Community                       & Limited                      & Occasional                  \\
\end{tabular}
\end{adjustbox}
\end{table}
\begin{table}
\caption{Definitions and count of categorical values describing the developers in Table~\ref{tab:descriptives}}
	\label{tab:casecount}
\begin{adjustbox}{width=\textwidth}
\begin{tabular}{ l | p{0.24\linewidth} | r | p{0.76\linewidth}}
    
    \textbf{Characteristic}                      & \textbf{Category}      & \textbf{Count}      & \textbf{Description} \\  \hline
    Experience Level     & Entry Level   &  3 & Developers working on reviews within academic projects or in the beginning of their career.\\ 
    \rowcolor[HTML]{EFEFEF} \cellcolor[HTML]{FFFFFF}                                                        & Mid-Career   & 14 &   Developers who have several years of experience, taking responsibility for parts of the code-base.\\ 
                                                            & Supervisory  &  5 &  Developers who apart of codebase take responsibility for people, leadership and management functions and provide mentoring during reviews.  \\  \hline
    \rowcolor[HTML]{EFEFEF} \cellcolor[HTML]{FFFFFF}Environment            & Internal        & 18 &  Mostly stable teams working on an internal project in a company (including teams working on an OSS code) with rather regular and face-to-face contact.\\ 
                                                            & Community  & 10 &   Organically formed open-source community. \\ 
    \rowcolor[HTML]{EFEFEF} \cellcolor[HTML]{FFFFFF}        & Academic (Researcher) &  5 &   Working in the academia as a researcher or teaching using code reviews;  \\  
                                                            & Academic (Student)   &  2 &  Doing code reviews for educational purposes or small individual projects while studying. \\  \hline
    CR Process Maturity & High   &  14 &  Code review is an established and regulated part of the software development cycle.\\ 
    \rowcolor[HTML]{EFEFEF} \cellcolor[HTML]{FFFFFF}    & Limited   &  6 &   Code review is a defined process but is not performed as an established part of a full development cycle (\eg educating students, performing reviews in a small scale company with limited reviewing options). \\ 
        & Ad-hoc  &  2 &  Reviewing code happens on an irregular basis, and the code review process is not well established. \\  \hline
    \rowcolor[HTML]{EFEFEF} \cellcolor[HTML]{FFFFFF}Conflict Frequency & Low   &  2 &  Not encountering conflicts or referring to not having much experience.\\ 
        & Occasional  &  9 &   Encountering conflicts `rarely' or `sometimes'.\\
    \rowcolor[HTML]{EFEFEF} \cellcolor[HTML]{FFFFFF}                                                        & Regular & 11 &  Generalized experience with conflicts or their repeated and regular presence in code review.\\

\end{tabular}
\end{adjustbox}
\end{table}

\subsection{Analysis}
\label{sec:met:analysis}
To explore the conflicts during code review in a broad perspective based on developers' experience, we used Thematic Analysis~\cite{clarke2015thematic} with a bottom-up approach. This approach is an inductive one where the researcher derives codes and themes from the data content~\cite{clarke2015thematic}. Unlike the top-down (deductive) method, where the researcher creates codes in advance (based on some concepts, topics, or ideas), in the bottom-up approach, the codes and themes that the researcher derives closely match the data content (Appendix \ref{sec:appendixV} provides further details).

Before and during the qualitative analysis, we referred to the literature about interpersonal conflicts, code review, and related areas to interpret the data in relation to the existing body of knowledge.
To analyze the data, we used a qualitative data analysis software named NVivo 12.\footnote{\url{https://qsrinternational.com/nvivo}}

Thematic analysis consists of six steps~\cite{clarke2015thematic}:
\begin{enumerate*}
  \item Familiarizing with the data,
  \item generating initial codes,
  \item searching for themes,
  \item reviewing potential themes,
  \item defining and naming themes, and
  \item writing a report.
\end{enumerate*}

Step 1 (\textit{familiarizing with the data}) was used to read through all the interviews and get to know the data. The goal was to create an initial mind map of the main issues and relationships in the data, clear out how to see the conceptual dimensions of interpersonal conflict as defined in the literature~\cite{hartwick2002conceptualizing}, and specify what type of data relates to each research question. Throughout the reading, the data was separated into three datasets in order to analyze the RQs individually. Dataset 1 included parts of the interviews that related to RQ1 and RQ2. Dataset 2 included references for RQ2 and RQ4, and Dataset 3 included references to RQ3 and RQ5.
The reason was that specifying differences in factors, focus, and disagreement of conflicts was only developing. The overlaps of datasets were resolved during later steps of the analysis.

Steps 2--6 were executed iteratively and separately for each dataset with the help of related researcher memos.
During the \textit{initial coding phase} (step 2), dataset overlaps were resolved. Individual pieces of information related to the research questions were formulated and coded in more abstract terms/codes. The initial structure of data started to emerge as well due to multiple research questions and differing levels of abstraction that participants used to express their experience.
In step 3 (\textit{searching for themes}), initial codes were grouped into themes on a higher level of abstraction. This phase was important for (dis)confirming emerging themes through finding (dis)similarities between codes. Important questions included: `What makes these narratives same/different?' `What would be different if this aspect of the narrative would change?' `Are these topics related/different/interacting?'

Step 4 (\textit{reviewing potential themes}) consisted of controlling for overlaps and duplication and going back to the raw data in cases when a decision or clarification was needed. We searched for opportunities for merging themes together to simplify the final data structure.
During this step in the last dataset, we conducted a control for overlaps of RQs and themes, simplified and unified the presentation of the results to avoid unnecessary complexity and content duplication.

Step 5 (\textit{defining and naming themes}) was done to provide and write final themes' definitions and names while analyzing the entire data and all available context.

While \textit{writing the report} (the paper) in step 6, we focused on presenting not only the resulting themes but also the context in which they are mentioned, and we aimed to convey the story present in the data while respecting both available knowledge and developers' narratives.

\subsection{Determining Validity}
We used three ways of validation of the analysis that reflect the context of the data collection. The interviews were collected only by one researcher who also did the analysis. Due to the size of the data, it was not possible to involve further people who would be deeply familiar with their content. The validation procedures are disconfirming evidence, peer debriefing, and an audit trail-like process~\cite{creswell2000determining}, further described in Appendix \ref{sec:appendixV}.

To \textit{(dis)confirm evidence}, attention was paid during the analysis to consider differences between narratives of different participants and situations. We explicitly searched for similarities that would confirm the importance of a theme or disconfirm the current interpretation of the data.

Throughout the analysis, the first author who conducted and analyzed the interviews was discussing the progress and process of the analysis and challenges with data interpretation with a colleague in the research group, who was acquainted with the content of the interviews, performing so-called \textit{peer debriefing}.

After the analysis, materials about the analysis (available in the supplementary material), methodology, and results were collected for conducting an \textit{audit-like process} by a researcher who was not involved directly in the analysis but who was involved in the study design and execution---the third author of the study. The goal was to examine both the process and product of the analysis and determine the trustworthiness of the findings. The researchers discovered no invalid findings through this process.

\subsection{Limitations}
Despite having followed the described research method rigorously, our study has the following limitations:

\begin{enumerate}
  \item The study is exploratory and focused on uncovering a broad context of conflicts during code review. Even though this process is sufficient to uncover a wide variety of factors, it cannot provide details of the process in which they intervene in the conflict dynamics.
  \item We adopt a data-driven perspective and collect the experience of developers with interpersonal conflicts during code review. We work with literature from software engineering and other fields throughout the analysis to maximize the utility of knowledge generated from developers' data. However, interpersonal conflicts (also in the organizational environment) are a knowledge-rich area, and there is a substantial body of knowledge that our study cannot fully incorporate. It is our hope that many theory-driven perspectives on this issue might offer other valuable insights.
 \item We aimed for code saturation, rather than meaning saturation. Aiming for meaning saturation would, in our case, lead to a study that would require a much greater dataset, with many more interviews, and extensive analysis. For example, while we could identify a rich set of factors, we could not describe all its ways of intervention in conflicts.
 \item We reach a diverse sample of interviewed developers. However, two developer profiles seem to be harder to recruit for such a study: stubborn and rigid developers and developers who support conflicts. The sample contains two developers who see conflicts positively, and one developer defines himself as a `perfectionist,' but our results are likely not able to cover these two missing profiles.
  \item The interviews are collected and analyzed only by the first author. To minimize subjectivity and ensure the analysis's validity, the first author performed peer debriefing with the second author, who had an overview of all the sample interviews. Moreover, the third author performed control on the audit trail.
\end{enumerate}

\section{Results}
Following the aforementioned method, we conducted 22 interviews, manually analyzed 167 pages of transcripts containing 145,496 words, and organized 5,655 individual codes into the themes presented in the rest of this section. When quoting developers, we use a [D$X$] notation, where $X$ is the developer's unique identifier reported in Table~\ref{tab:descriptives}.

\subsection{\rqZero}
\label{sec:rq0}
Developers report a sense of normality around conflicts during code review, and all but one interviewee have experienced conflicts during code review at least once.

Developers perceive conflicts as a \textbf{normal} aspect of the code review process (D1, D10, D11, D13, D15, D17, D19, D21, D22).
Code review conflicts are also \textbf{specific}. While some developers do not distinguish conflicts in code review context from other conflicts in the workplace, they also report that conflicts are \textit{easier to happen} during a code review (D9, D18) or it is the \textit{only place for conflicts} (D18) at work. Furthermore, code review conflicts \textit{directly affect developers' work} (D4, D6, D17, D20) (\eg developers' place in a team, the progress of their work). Code review comprises feedback on, evaluation, and judgment of the developers' work assessed by colleagues. Developers report that code review conflicts tend to get \textit{personal} (D12, D16) and happen in a \textit{specialized context} - discussing code-related decisions with fellow developers (D2, D5, D6, D9, D11, D18, D22).
Discussions of one's work and abilities, which might be personally sensitive, occur through asynchronous communication (\eg code review tools, emails, task managers) that does not primarily allow for personal connection. Code review conflicts are dependent on the relations among individuals who participate in the code review, but they are focused on the code. To summarize, code review conflicts are \textit{non-technical conflicts about technical issues}. As one developer put it: \qud{Usually, the problems are not technical. That's what I learned while working. \ldots~If somebody else is doing [the] research and that thing is really technical and difficult, usually the problem is interacting with that person even if it's about the code}{15}.

Developers also report that conflicts are hard to escape and \textbf{hard to anticipate or handle} (D8, D9, D18), and when conflicts happen, developers are unsure of what to do. An interviewee explained: \qud{I don't have a lot of conflicts during code reviews. And when I do it's usually \ldots~I usually don't know what to do}{8}

In addition, developers find code review conflicts \textbf{unpleasant but potentially beneficial} (D6, D8, D12, D14, D15, D17, D22). Conflicts are related to negative emotions. However, conflicts can be constructive and sometimes necessary to find a resolution and meet the code review goals. For example: \qud{You can have a code review with someone senior that doesn't want conflict, and he doesn't do a proper code review. There's no tension at all, but also there's no improvement, so the goal of the code review is not met}{15}.

\subsection{\rqOne}
To answer RQ2, we decompose and analyze code review conflicts referring to the two-dimensional construct by~\citet{hartwick2002conceptualizing}. This construct comprises \textit{conflict focus} and \textit{conflict components} (\ie \textit{disagreements}, \textit{interference}, \textit{emotions}). Table~\ref{tab:components} summarizes the themes and sub-themes that emerged for conflicts' focus and components.

\begin{table*}
  \caption{RQ2: Conflicts Focus and Components in Code Review}
  \label{tab:components}
  \begin{adjustbox}{width = \textwidth}
    \begin{tabular}{ll|ll}

 \textbf{Dimension}                      & \textbf{Participant ID}           & \textbf{Dimension}             &   \textbf{Participant ID}               \\ \hline
\multicolumn{2}{c|}{\textbf{CONFLICT FOCUS}}                 & \multicolumn{2}{c}{\textbf{CONFLICT COMPONENTS}} \\ \hline
\textit{Functional   Aspects}           &                                            & \multicolumn{2}{l}{\textbf{Interference}}                                \\ \hline
Features                                & D8                                         & \textit{Public}                & \textit{}                               \\ \cline{3-4}
\cellcolor[HTML]{EFEFEF}Incomplete functionality                & \cellcolor[HTML]{EFEFEF}D2, D12                                    & Open argument                  & D2, D5, D13, D14, D17, D2              \\ \cline{1-2}
\textit{Non-Functional   Aspects}       &                                            & \cellcolor[HTML]{EFEFEF}Ignoring each other            & \cellcolor[HTML]{EFEFEF}D1, D6, D11, D12, D19, D21              \\ \cline{1-2}
Design/Big issues                       & D11, D13, D15, D22                         & Forcing actions                & D5, D7, D11, D12, D21                   \\
\cellcolor[HTML]{EFEFEF}Security                                & \cellcolor[HTML]{EFEFEF}D11, D12                                   & \cellcolor[HTML]{EFEFEF}Blocking actions               & \cellcolor[HTML]{EFEFEF}D5, D6, D8                              \\ \cline{3-4}
Performance                             & D1, D6, D11, D19                           & \textit{Hidden}                & \textit{}                               \\ \cline{3-4}
\cellcolor[HTML]{EFEFEF}Implementation                          & \cellcolor[HTML]{EFEFEF}D6, D8, D9, D15                            & Lack of progress               & D1, D5-D9, D11, D14                     \\
Maintainability and   Readability       & D1, D2, D6, D7, D11-D15, D18, D19, D21     & \cellcolor[HTML]{EFEFEF}Proactivity                    & \cellcolor[HTML]{EFEFEF}D1, D6-D9, D13-D19, D21                 \\ \cline{1-2}
\textit{Bad procedures   and practices} & \textit{}                                  & Feeling the emotions           & D6, D7, D22                             \\ \hline
Conventions adherence                   & D4, D6, D7, D10, D12, D15, D21             & \textbf{Emotions}              &                                         \\ \cline{3-4}
\cellcolor[HTML]{EFEFEF}Unclear requirements                    & \cellcolor[HTML]{EFEFEF}D12                                        & \textit{Basic}                 &                                         \\ \cline{3-4}
Lack of care about the review   process & D6, D22                                    & Fear                           & D4                                      \\ \cline{1-2}
\multicolumn{2}{c|}{\textbf{CONFLICT COMPONENTS}}            & \cellcolor[HTML]{EFEFEF}Sadness                        & \cellcolor[HTML]{EFEFEF}D2, D15                                 \\ \cline{1-2}
\textbf{Disagreement}                   &                                            & Anger                          & D2, D7, D8, D12-D14, D18, D19           \\ \hline
\textit{Task definition}                & \textit{}                                  & \textit{Higher}                &                                         \\ \hline
Definition of done                      & D5, D6, D8, D9, D11, D12, D19, D21, D22    & Anxiety                        & D7                                      \\
\cellcolor[HTML]{EFEFEF}Goal                                    & \cellcolor[HTML]{EFEFEF}D2, D6, D7, D9, D12, D19                   & \cellcolor[HTML]{EFEFEF}Frustration                    & \cellcolor[HTML]{EFEFEF}D5-D7, D11, D18, D20, D21               \\
Priorities                              & D2, D4, D6, D8-D10, D12, D13, D16, D18-D21 & Hate                           & D12                                     \\ \cline{1-2}
\textit{Quality assessment}             & \textit{}                                  & \cellcolor[HTML]{EFEFEF}Shame                          &  \cellcolor[HTML]{EFEFEF}D7                                      \\ \cline{1-2}
Code quality                            & D1, D2, D5-D9, D11, D12, D16, D19, D21     & Boredom                        & D21                                     \\
\cellcolor[HTML]{EFEFEF}Review quality                          & \cellcolor[HTML]{EFEFEF}D5-D7, D9, D12, D18, D21                   & \cellcolor[HTML]{EFEFEF}Disappointment                 & \cellcolor[HTML]{EFEFEF}D1, D6, D22                             \\
Assessment perspective                  & D1, D2, D5-D9, D11, D12, D16, D19, D21     & Unhappiness                    & D6                                      \\ \cline{1-2}
\textit{Information availability}       & \textit{}                                  & \cellcolor[HTML]{EFEFEF}Worry                          & \cellcolor[HTML]{EFEFEF}D4, D17                                 \\ \cline{1-2}
Work constraints                        & D4, D6, D13-D15, D22                       & Discomfort                     & D4, D6, D7, D11, D14                  \\
\cellcolor[HTML]{EFEFEF}Knowledge and information   usage       & \cellcolor[HTML]{EFEFEF}D1, D5-D7, D11, D12, D16-D19, D21          & \cellcolor[HTML]{EFEFEF}Awkwardness                    & \cellcolor[HTML]{EFEFEF}D11                                     \\ \cline{1-2}
\textit{Social perception}              & \textit{}                                  &                                &                                         \\ \cline{1-2}
Acceptable behavior                     & D1, D17, D18, D22                          & \textit{}                      & \textit{}                               \\
\cellcolor[HTML]{EFEFEF}Comparison                              & \cellcolor[HTML]{EFEFEF}D1, D11, D14, D15, D17                     &                                &                                         \\
Interpretation                          & D1, D6, D7, D9, D15, D17, D18              &                                &                                         \\
\end{tabular}
\end{adjustbox}
\end{table*}

\subsubsection{Focus of the Conflict}
\label{sec:conflict_focus}
This point addresses what the conflicts during code review are about. Our results can be summarized under the following areas:

\begin{description}[leftmargin=0.3cm]

  \item[\textbf{Functional aspects.}]
  Conflicts on functional aspects comprise what features and how they should be implemented and the completeness of the features' functionality.

  \item[\textbf{Non-functional aspects.}]
  Conflicts on non-functional aspects can be about solutions to optimize code performance, security, and maintainability. Conflicts on non-functional aspects can also be about how the code change is implemented and how the code change fits the overall software design. Our findings also indicate that a significant amount of conflicts on non-functional aspects regard code readability (\ie code style, adherence to the code style conventions, consistency of the code style throughout the code base).

  One interviewed developer explained: \qud{Some people \ldots~take their personal taste on code like if it was of the whole company. If they use two spaces instead of four, it's like you will see like 50 comments on the code review like <<delete to a space, delete to a space, delete to a space>>. For them, it's wrong for you it's not. You cannot make a discussion on it because it doesn't have positive or negative points. It's just two spaces or four spaces. It's nothing. It's code style. And they take it like super personal, and they impose on you their personal style. This was a complicated thing also to deal with because it has actually a very easy fix which was to impose general code style \ldots~ but some people actually didn't like it as well}{18}.

  \item[\textbf{Bad procedures.}]
  Conflicts on inadequate procedures emerge about following standards, guidelines, and conventions on the overall software development process, including code implementation, code reviews, and documentation. One developer explained: \qud{What pisses many people off is: there are guidelines on how to open a code review and how to write things, how to write the problems, so before opening issues or before opening pull-requests, etc. have a look at those guidelines...}{4}.

\end{description}

\subsubsection{Disagreement}
Disagreement is the cognitive component of conflicts referring to the divergence of conflict participants' values, needs, interests, opinions, goals, or objectives. We identified disagreements in the following areas:

\begin{description}[leftmargin=0.3cm]

  \item[\textbf{Task definition.}]
  Reported disagreements on task definition are due to misalignments in understanding the \textit{goal} of the review/change, what is (not) part of the task, the \textit{definition of done}, or who is responsible for the task.
Developers have different thresholds when a change in the code is needed or where is the balance between optimal and sufficient code: \qud{Sometimes it's just like if I am not following the principles of a programming language or something, but it's not relevant because the code is working and it's secure, and it's functional, but if I am not following the rules of the coding language it doesn't matter. It's not relevant for the project, and I feel like it's mean. It is not relevant for the code review}{12}.

One main factor related to disagreements on task definition is developers' \textit{prioritization} of what to include in the code implementation or feedback. For instance, a developer might prioritize fixing an issue that might seem irrelevant to the other.

  \item[\textbf{Quality assessment.}]
  Differences in the perception of the \textit{code quality} can spin a conflict; also, the perception of what is \textit{useful and relevant} in a code review might differ. As one developer put it: \qud{When I was writing the code, I already included in it all the things she was mentioning, and it seemed to me totally irrelevant. To the extent that I felt like the comments of the people are there for nothing}{6}.

It can be problematic when developers view the code from a different \textit{perspective} (\eg as code author or reviewer). There can be misalignment in assessing the code complexity to estimate the effort and resources needed for the code's implementation or review. Problems can also occur due to differences in the abstraction levels. For instance, a developer might base the discussions on code-level without an insight into the overall software design or vice versa.

  \item[\textbf{Information availability.}]
  Our findings also indicate disagreements based on differences in the information available to each conflict party.
  Developers use different resources to obtain \textit{knowledge}, which might cause issues during code review when those information resources disagree. Experience-based knowledge might be incompatible with the knowledge developers gain by studying current practices and state-of-the-art and can also be outdated. Besides, misunderstandings can emerge when knowledge about the software system and architecture or procedures agreed upon within the project team or organization are not available to all developers.

  Developers also might lack mutual understanding of their colleagues' \textit{work constraints}. These limit the developer's ability to react to requested changes. Developers report struggling with having sufficient time, energy, or other resources to deal with potential problems. A reviewer can make requests that are infeasible or not desirable for the developer simply because the reviewer is either unaware or inconsiderate of the developer's work constraints.

  One developer explained: \qud{Imagine I work on a feature. I am late. I've been working for three or four days, my boss is putting pressure on me to deliver and here comes [name of person] and gives me feedback that will require two days of work. You are not going to accept it very well}{15}.

  \item[\textbf{Social perception.}]
  Social perception refers to inferring others' intentions, thoughts, or personality traits based on our perception of their observable behaviors (\eg verbal or non-verbal communication)~\cite{dijksterhuis2001perception}. However, the process is prone to errors and misinterpretations~\cite{funder1987errors}.
  One of the \textit{(mis)interpretations} developers mentioned in the interviews is whether one views the feedback as assessing the quality of the code or the developer who implemented the code. The perception of the latter can make the debate more personal. Furthermore, developers can interpret the intention to help identify their insufficiency to solve the problem at hand.
  \qud{And the other person (not necessarily) wants to fix it or give feedback so that it is better, but I perceived it in the moment like it is not good enough. And that can not hurt you but it tells you: <<Well, you could have done it in the first attempt.>>}{6}.
  
  Some developers can \textit{compare} themselves to others. Developers also might compete in who is right or who is the better programmer.
  Furthermore, each individual has a different perception of \textit{acceptable behavior}, which might lead to conflicts. Like so, developers step over what others consider an unacceptable interference with their actions and intentions: \qud{Then they say, <<I did not like this, I changed it, I sent it.>> Which I find strange. That's too much when somebody touches your work. It's really not nice}{1}.

\end{description}

\subsubsection{Interference}
Interference is the behavioral component of conflicts, which comprises conflict parties' actions and how the conflict can manifest. We report different interference types during code review based on the conflicts' categorization as \textbf{public} and \textbf{hidden}, referring to the literature~\cite{kolb1992hidden, Dubinskas:1992, Kolb:1992, Bartunek:1992} we previously mentioned (Section~\ref{sec:related_work_general}). Public conflicts include behaviors directly interfering with the attainment of one's goals, whereas hidden conflicts are not expressed directly.

\begin{description}[leftmargin=0.3cm]

  \item[\textbf{Public conflicts.}]
  We identified four types of behavior during a code review that can interfere with the attainment of the developer's goals: \textit{heated argument}, \textit{blocking}, \textit{forcing}, and \textit{ignoring}.
  Developers can \textit{block} other developers from achieving their goals. For instance, blocking can be in the form of pushback on change acceptance, as also reported by \citet{egelman2020pushback}, or simply not allowing a developer to make changes that the developer sees as necessary: \qud{I want to remove some \ldots~blocks of code and they don't let me do it}{5}.
  
  \textit{Forcing} developers to do something they do not find correct is also another type of such interference: \qud{When they force me to do something wrong it's very rare now, but sometimes I do it, I do it as fast as I can so I can jump again for the right way}{5}.
  
  Our findings also indicate \textit{ignoring} and disregarding the opinions and requests of the counter-party as other interference.

  \item[\textbf{Hidden conflicts.}]
  In this category, we present a list of symptoms that are characteristic of conflicts. Furthermore, these symptoms of conflicts can also be present in public conflicts. Developers report that, in some situations, there are no obvious signs of a conflict in the communication or behaviors, but the participants \textit{can detect the emotions}. One developer explained: \qud{\ldots~when I notified her about it few times, and it was still happening to her, that she did not pay attention, I could see she is not happy with that. But I cannot say she said: <<No, I will not do it because I don't want.>> \ldots~In terms of the comment, you cannot really tell}{6}.
  
  Hidden conflicts during code review are accompanied by \textit{proactivity} towards finding a solution. Developers use various strategies like reminding others what should be done or sneaking into code reviews they do not need to be part of to have more control over the codebase. Continuing the discussion outside the code review tool using face-to-face communication, involving other colleagues or a manager, or addressing the issue during a team meeting can all be a symptom of a conflict. Another indication of proactivity during code review is writing long and explanatory comments. Conflicts can also be accompanied by a `ping pong' of comments resulting in a long discussion \textit{lacking progress}: \qud{We talk about it for an hour, and nothing is discussed, nothing will happen, and everyone is still in their stance}{1}.

\end{description}

\subsubsection{Emotions}
Emotional response to a stimulus (\eg a discrepancy between needs and goals) is a natural process. Table~\ref{tab:components} lists the negative emotions developers reported during the interviews. Emotions can be categorized as basic and higher~\cite{clark2010relations}. \textbf{Basic emotions} (\eg fear, anger, sadness) can be activated by unconditioned stimuli and lead to automatic, involuntary reactions. A learning process determines \textbf{higher emotions}. For instance, people learn what situations make them frustrated or ashamed. Therefore, experiencing higher emotions can also be altered by the learning process~\cite{clark2010relations}.

Emotions that arise during code review include higher emotions alongside basic emotions since code review requires higher cognitive processes such as reasoning, analysis, and decision-making and is a socially complex situation. As a result, although negative emotions during code review cannot be avoided entirely, developers can learn new ways to interpret the situations resolving higher negative emotions. For instance, developers can learn to accept feedback, give constructive feedback, and learn about their peers' communication styles.
Moreover, understanding negative emotions can provide information about what needs to change to mitigate them. For instance, frustration appears when a person encounters obstacles in a striving path towards their goals~\cite{berkowitz1989frustration}. On the other hand, anxiety can yield as one might not know what to do in a specific situation or feels insecure about his abilities to meet particular demands~\cite{keedwell1996anxiety}.

\subsection{\rqTwo}
\label{sec:RQ3}
Our findings indicate that conflicts impact code reviews, the \textbf{code}, the \textbf{developer} being part of the review, the \textbf{team} developer interacts with, and the \textbf{organization} the developer works for. These four are nested layers that are related to each other. To illustrate how a conflict's consequence on one layer can impact other layers, we firstly present the following scenario for conflicts' negative consequences based on themes and relationships (also in Table \ref{tab:consequences}) we identified in the analysis:

\begin{table*}
  \centering
  \caption{RQ3: Positive and Negative Consequences of Conflicts}
  \label{tab:consequences}
  \begin{adjustbox}{width=\textwidth}
\begin{tabular}{l|p{4.5cm}p{3cm}|p{4.5cm}p{3cm}}
                                  & \textbf{Negative}                                               &  \textbf{Participant ID}                          & \textbf{Positive}                    				&  \textbf{Participant ID}               \\ \hline
\rowcolor[HTML]{EFEFEF} \cellcolor[HTML]{FFFFFF} \textbf{Code}             & Lower code   quality                                           &  D5, D8-D11, D21                                & Better code quality                			&  D2, D4-D6, D8                         \\
\textbf{}                     & Worse   maintainability                                       &  D11, D21                                            & Better maintainability            				&  D11                                   \\
\rowcolor[HTML]{EFEFEF} \cellcolor[HTML]{FFFFFF} \textbf{}                     & Following bad   practices                                    & D5                                                       & Standardization                                 & D1, D2, D5, D7, D18                   \\ \hline
\textbf{Developer}      & Loosing trust   in the review process                  & D1, D4, D6, D7                                    & Initiating changes                              & D2                                    \\
\rowcolor[HTML]{EFEFEF} \cellcolor[HTML]{FFFFFF} \textbf{}                     & Disengagement                                                  & D1, D4, D6-D8, D12, D15                    & Generating new ideas and solutions  & D17, D19                              \\
\textbf{}                     & Lower   productivity                                            & D1, D5, D11, D17, D21                        & Getting own changes approved         & D5                                    \\
\rowcolor[HTML]{EFEFEF} \cellcolor[HTML]{FFFFFF} \textbf{}                     & Leaving/Loosing   job                                         & D4, D11, D17, D21, D22                      & Professional growth                          & D1, D2, D4, D5,. D10, D11, D13,   D16 \\
\textbf{}                     & Reviewer/Reviewee   Selectivity                           & D1, D8, D9, D15, D18, D19                  & Improved relationship towards work & D5, D8, D14                           \\
\rowcolor[HTML]{EFEFEF} \cellcolor[HTML]{FFFFFF} \textbf{}                 	   & Negative   relationship towards work                  & D1, D2, D5-D7, D9, D10, D12, D15     & Positive emotions                              & D2, D5, D17                           \\
\textbf{}                     & Negative   emotions                                            & D1, D2, D6, D7, D12, D17, D18           &                                       					 &                                       \\ \hline
\rowcolor[HTML]{EFEFEF} \cellcolor[HTML]{FFFFFF} \textbf{Team}            & Impaired   collaboration                                      & D1, D2, D5, D6, D8, D9, D11-D18       & Better collaboration                           & D11, D14-D17                          \\
\textbf{}                     & More conflicts                                                     & D8, D18, D21                                       & Better communication                        & D4-D7, D14-D16, D18, D21              \\ \hline
\rowcolor[HTML]{EFEFEF} \cellcolor[HTML]{FFFFFF} \textbf{Organization} & Bad reviews                                                       & D17                                                       & Better rentability                                 & D5                                    \\
\textbf{}                     & Delays                                                               & D4, D6, D9, D10, D21                            & Better working conditions                   & D5                                    \\
\rowcolor[HTML]{EFEFEF} \cellcolor[HTML]{FFFFFF} \textbf{}                     & Financial losses   from underutilized workforce and bad practices & D5, D12                          &                                      &                                       \\ 
\end{tabular}
\end{adjustbox}
\end{table*}

\begin{description}[leftmargin=0.3cm]

  \item[\textbf{Code.}]
  A team experiences conflicts on poor coding practices. Due to the struggle to improve them, the team produces code with lower quality and maintainability issues. Moreover, developers keep following poor coding practices. One developer explained: \qud{They just bypass me go to the team leader for example and okay <<it's working, yeah, go>> and this contaminates the code basically}{21}.

  \item[\textbf{Developer.}]
  Due to conflicts, developers report to lose trust in the code review process and disengage from contributing to code review discussions or the project. Disrupted code review discussions lead to poor decisions followed by increased developers' stress and negative emotions. Developers' motivation and productivity deteriorates due to the conflict and loss of trust in the fairness and efficiency of the code review process (\eg no trust in the feedback received or in colleagues). As one developer put it: \qud{I think that your views of the other members of the team change. That you get an opinion like, <<That guy's a bit weird>> or <<He thinks too much about himself,>> or <<He thinks that everything he does is good,>> and then there is this grumpiness in the team. When it comes to solving something, you do not want to deal with that person because you already feel he will criticize you strangely. That you do not trust the review anymore, as it should be}{1}.

  \item[\textbf{Team.}]
  The conflict leads to further issues with collaboration in the team. Developers start to have negative impressions about their colleagues, their relationships get colder, and the team creates a hostile environment which breeds further conflicts: \qud{I think that impacts a bit in other conflicts because if I have a conflict with someone in the code, this conflict \ldots~I will bring in the other conflict too, so I think it impacts on the other conflict and more conflicts. At some point you just go out and get crazy and start fighting.}{21}.

  \item[\textbf{Organization.}]
  As a result, the organization might suffer financial losses from an underutilized workforce, following lousy coding practices, higher turnover of employees and contributors, delays in deployment, or losing clients.

\end{description}

As mentioned in Section \ref{sec:rq0}, developers report that conflicts during code reviews can be unpleasant yet beneficial. Developers also report the importance of resolving conflicts and turning them into an opportunity to learn or improve. Therefore, to illustrate how a conflict's consequence in one layer (\eg code) can impact other layers (\eg developer, team, organization), we also present the following scenario for conflicts' positive consequences based on themes and relationships we identified through data analysis:

\begin{description}[leftmargin=0.3cm]

  \item[\textbf{Code.}]
  Developers resolve the conflicts that arise due to poor coding practices by the collaborative formulation of new coding conventions in the team. Consequently, code quality and maintainability improve. Through the confrontation of ideas, new creative solutions are found.

  \item[\textbf{Developer.}]
  Developers learn from each other and their own mistakes, leading to improved technical skills (\eg coding) and knowledge in addition to improved communication and decision-making skills. One interviewee explained: \qud{I can realize that I am doing something wrong. I can learn from it, and it will be better for everybody. It will be better for me, that I will make it better, and it will be better for the application and it will be better for the other developers who will come and work on it because it will be better planned, better fixed and so on}{11}.

  \item[\textbf{Team.}]
  Successful handling of the conflict results in more fruitful discussions and better collaboration within the team. Developers improve how they communicate with each other, their mutual understanding, learn about others' opinions, and accept criticism: \qud{From personal experience [with] code reviews I think if you have tension and you solve the issues you are in a better place than when you started because people understand you better and they are more comfortable with you. If you let things escalate and you don't talk, and then you want that person out of work. It's very complicated}{15}.

  \item[\textbf{Organization.}]
  The organization profits from better use of their investments in their workforce and better working conditions that arise from the discussing practices followed in the organization.

\end{description}

\subsection{\rqThree}
\label{sec:factors}
Factors are the characteristics of the environment that play a role in the presence and severity of conflicts---they are predictors, moderators, or possible areas of prevention for conflicts. Table~\ref{tab:factors} presents a summary of the factors developers mentioned during the interviews. Similar to explaining conflicts' consequences (Section~\ref{sec:RQ3}), we explain factors under the following categories:

\begin{table*}
  \caption{RQ4: Factors Involved in Conflicts during Code Review}
  \label{tab:factors}
  \begin{adjustbox}{width=\textwidth}
\begin{tabular}{lp{3.5cm}|lp{3.5cm}}

\textbf{Code and   Review}                       &  \textbf{Participant ID}                                         & \textbf{Developer}                                 &  \textbf{Participant ID}     \\ \hline
\textit{Change}                                        &  \textit{}                                                                                         & Awareness of behaviour in the conflict   &  D6, D14, D15, D17, D19 \\ \cline{1-2}
Challenging review                                   &  D1, D6, D7, D9, D11, D12, D14, D18-D20         & \cellcolor[HTML]{EFEFEF}Engagement                                           &  \cellcolor[HTML]{EFEFEF}D1, D2, D5-D15, D17-D21 \\
\cellcolor[HTML]{EFEFEF} Change importance                                  &  \cellcolor[HTML]{EFEFEF}D1, D2, D7, D11, D13, D15-D17, D19               & Personality                                              &  D1, D2, D4-D15, D17-D19, D21, D22 \\
Quality of the change code and   solution &  D6, D9, D12, D22                                              & \cellcolor[HTML]{EFEFEF}Profiting from the review                         &  \cellcolor[HTML]{EFEFEF}D1, D2, D5, D9, D13-D16, D20-D22     \\
\cellcolor[HTML]{EFEFEF}Review size                                               &  \cellcolor[HTML]{EFEFEF}D1, D11, D15, D16, D18-D20                            & Developer role                                        &   D6                                                   \\
Type of code                                             & D22                                                                    & \cellcolor[HTML]{EFEFEF}Review role                                              &  \cellcolor[HTML]{EFEFEF}D1, D8, D11, D14, D20, D22    \\ \cline{1-2}
\textit{Codebase}                                      & \textit{}                                                                                         & Psychological state                                  &   D1, D2, D5, D7, D15, D18, D22  \\ \cline{1-2}
Code quality                                              &  D1, D5, D6, D9, D11-D13, D15, D16, D19, D21 & \cellcolor[HTML]{EFEFEF}Workload                                                &   \cellcolor[HTML]{EFEFEF}D1, D2, D6-D9, D11-D18, D21, D22  \\ \cline{3-4}
\cellcolor[HTML]{EFEFEF}Documentation quality                              &  \cellcolor[HTML]{EFEFEF}D6, D16                                                              & \textit{Experience}                                  &                                                      \\ \cline{3-4}
System complexity                                    &   D7, D19, D20                                                     & Hierarchy                                                &  D3, D6, D7, D9-D11, D14-D19, D21, D22      \\ \cline{1-2}
\textit{Review policy}                                & \textit{}                                                                                           & \cellcolor[HTML]{EFEFEF}Learning curve                                       &  \cellcolor[HTML]{EFEFEF}D1, D4, D6-D10, D12-D15, D17, D1, D20-D22            \\ \cline{1-2}
Flexibility                                                  &  D1, D6-D10, D18                                                 & Priorities                                               &  D6, D10, D11                                         \\
\cellcolor[HTML]{EFEFEF}Frequency of reviews                                &  \cellcolor[HTML]{EFEFEF}D6, D7, D10                                                         & \cellcolor[HTML]{EFEFEF}Reviewer experience                              &  \cellcolor[HTML]{EFEFEF}D1, D4, D6-D10, D14, D15, D20                        \\
Systematic reviewing                                &  D2                                                                        & Self-confidence                                     &  D1, D4-D7, D9, D10, D12, D14, D17-D19, D21   \\
\cellcolor[HTML]{EFEFEF}Value of review work                                &  \cellcolor[HTML]{EFEFEF}D7, D16, D22                                                        &                                                             &                                                      \\ \cline{1-2}
\textit{Review process}                             & \textit{}                                    															&	                                        					 &                                                      \\ \cline{1-2}
Binary output                             				  &  D7, D14, D16							                                &  						                                     &                                                      \\
\cellcolor[HTML]{EFEFEF}Availability and stability of   reviewers      &  \cellcolor[HTML]{EFEFEF}D1, D2, D6, D7, D10, D11, D16, D17                     &                                                             &                                                      \\
Solutions on par                                       &  D6, D7, D16, D19                                                  &                                                             &                                                      \\
\cellcolor[HTML]{EFEFEF}Lack of information on how to   decide    &  \cellcolor[HTML]{EFEFEF}D1, D6, D7, D11, D13, D16, D18                            &                                                             &                                                      \\
Subjectivity of decisions                           &  D11, D13-D16, D18, D19, D21                              &                                                             &                                                      \\ \hline
\textbf{Team}                                          & \textbf{}                                   															 & \textbf{Organization}                           &                                                      \\ \hline
Dependency on others                             &  D5-D10, D17, D18, D20                       				     & Culture                                                &  D1, D3-D7, D9, D10, D15, D21                         \\
\cellcolor[HTML]{EFEFEF}Distance and Diversity                             &  \cellcolor[HTML]{EFEFEF}D1, D2, D4-D6, D8-D11, D13, D14, D16-D21       & \cellcolor[HTML]{EFEFEF}Development process maturity             &  \cellcolor[HTML]{EFEFEF}D1, D2, D4, D7, D9, D11, D13, D14, D16, D17, D19-D21 \\
Majority                                                   &  D1, D5, D8, D14, D15, D18, D21                           & Leadership quality                                &  D5, D6, D9, D11                                      \\
\cellcolor[HTML]{EFEFEF}Size                                                         &  \cellcolor[HTML]{EFEFEF}D7, D18                                                                 & \cellcolor[HTML]{EFEFEF}Status                                                   &  \cellcolor[HTML]{EFEFEF}D2, D4, D6, D10, D11, D13, D17                      \\
Team composition                                   &  D9, D11, D14, D18, D20, D21                               & Size                                                       &  D6, D8, D9, D11, D21                                 \\ \hline
\textit{Communication}                           &                                                                                                          & \textit{Environment}                              &                                                      \\ \hline
Constructive feedback                             &  D1-D9, D11-D19, D21                                          & Internal                                                 &  D8, D11, D13, D14, D16, D21, D22                     \\
\cellcolor[HTML]{EFEFEF}Mode                                                      &  \cellcolor[HTML]{EFEFEF}D1, D4, D6, D7, D10, D17, D18, D22                      & \cellcolor[HTML]{EFEFEF}Community                                          &  \cellcolor[HTML]{EFEFEF}D8, D11, D13, D15, D16, D21                          \\
Misunderstanding                                   &  D1, D2, D4, D6, D8, D12, D15, D17, D18, D22       & Academic                                              &  D12, D13, D19, D20                                   \\ \cline{1-2}
\textit{Cooperation}                                & \textit{}                                                                                              &                                                              &                                                      \\ \cline{1-2}
Agreement on ways of   communicating  &  D14, D15, D17, D18                                               &                                                              &                                                      \\
\cellcolor[HTML]{EFEFEF}Common interest                                    &  \cellcolor[HTML]{EFEFEF}D11, D17, D18, D20                                               &                                                              &                                                      \\
Relationship quality                                &  D1-D4, D6-D9, D11, D13-D19, D22                       &                                                              &                                                      \\ 
\end{tabular}
\end{adjustbox}
\end{table*}

\begin{description}[leftmargin=0.3cm]

  \item[\textbf{Code and review.}]
  High code complexity and low code quality make it challenging for developers to solve issues efficiently, building tension and frustration. Reviews challenging due to \textit{code quality}, \textit{documentation quality}, \textit{review size}, including documentation and patch size and \textit{system complexity} also create a challenge to understand and communicate the issues in a code change and to propose optimal solutions. One developer said: \qud{This could be a problem if you have super challenging code reviews and you know you can do it, but it costs you a lot of time. I mean this can be a stress factor as well especially linking it to workload}{7}.

  Depending on the \textit{patch priority} and \textit{severity of the issue}, developers determine whether it is worth entering a disagreement: \qud{Well, if more people are discussing it, one can leave without a problem. But if he is one of those people who wrote or reviewed it, he could attend the discussion. And if it's something serious, like a bug, I would not do anything like that}{1}.

  \textit{Flexibility} to delegate the review and choose reviewers determines whether developers get stuck in a potentially unpleasant situation.  Code reviews occurring with high \textit{frequency} and intensity of contact among developers can lead to tension. Moreover, \textit{lack of value} put on code reviews either by individuals or by a team and company can result in poor software development (\eg coding, code review) practices and consequential tension and conflicts. One interviewee recounted: \qud{one developer \ldots~was known for not caring about the code reviews. That he always writes it somehow and then sometimes even does not do pull-request on GitHub and just sends the commit or he makes the PR, but he merges it by himself. So there was a discussion about either we have some rules, or we don't. And somebody always repeated this: <<Let's do it properly>> and \textit{(name)} just did it in his way again}{10}.

  The \textit{binary output} of a review (\ie `Accept' or `Reject') can be too simplistic and harder to accept. Furthermore, the \textit{availability} and \textit{stability} (\ie low turnover) of reviewers who could help clear up the situation or bring additional information facilitate decision-making during code reviews.

  \item[\textbf{Developer.}]
  Different \textit{developer roles} are related to different \textbf{code and review} characteristics thus making conflicts more or less likely to happen. For instance, full-stack developers might engage in code reviews more frequently and systematically than testers. Developers also have their own motivations during code review. \textit{Being aware} of differences in developers' roles and motivations or one's own mistakes is a step towards the resolution and management of conflicts during code reviews.
  Developers' priorities might differ depending on their \textit{role} as the author or reviewer of the code change (\eg to get things done vs. to optimize the code). The authors might also feel an attachment to their code and a knowledge of the rationale and process behind the code's implementation, which reviewers might not always have. The \textit{engagement} in one's work (\eg emotional attachment to the code as author) is crucial for active participation in the discussions but also may make developers more sensitive to negative feedback.

  Developers who experience \textit{autonomy}, have a feeling of accomplishment, and room for freedom are more engaged to spend  \textit{effort in doing a good job}, follow good practices, implement high-quality code, listen to others, and have productive discussions during code reviews. To deal with a conflict, it is also helpful to \textit{see profits} of the conflict (\eg learning something new, seeing the working result): \qud{Of course everyone can have conflicts and if the guy explained to me that I was doing something wrong and he explained to me in the way that I understand it \ldots~it's a positive conflict. Otherwise, it's useless.}{21}

  \textit{Psychological state} (\eg mood, fatigue, stress) can lower developers' ability to deal with negative feedback or the need to rework some changes. \textit{Workload} and deadlines contribute to the work-related stress and consequential behavior in the review discussion.

  Developers' \textit{personality traits} such as openness to change, resilience, perfectionism, sociability, emotionality, individuality, and conscientiousness affect how they engage in the review, discussion, and conflicts: \qud{If someone is stubborn, what can a man do?}{1}.

  Developers' \textit{experience} introduces hierarchy - more experienced developers having more authority and responsibilities. Developers report that novices need to build up self-confidence and improve coding skills and knowledge. Novice developers also need to gain experience in communicating their decisions and opinions about code and the code's implications for the project and handling different types of people, and how to get them on the same page. One developer explained: \qud{I accept that the reviewers have more experience than me. So I take the advice. There can be a conflict in the moment when both of the people are confident that their approach is good and the other person sees it differently.}{10}
  However, experienced developers might become rigid, slow to change, and over-confident in their knowledge. The novice developers can offer an up-to-date perspective and knowledge that can get lost due to the authority and unwillingness of experienced developers to adjust or novices' inability to discuss with experienced developers and confront them.
  Conflicts arise between experienced developers as well, but rather in the form of expert competition.

  \item[\textbf{Team.}]
  \textit{Distance and diversity} in a team (\eg geographical, cultural, personal, professional, organizational) can bring varied perspectives and create challenges in finding common ground and communicating clearly.
  Developers mention that team members should have good relationships, agreement on ways of communicating, and shared interest (\eg shared focus, shared code ownership, team effort towards a common goal). However, shared goals and good relationships do not always lead to constructive conflict resolution. For instance, the following may result in non-constructive conflicts: unproductive team habits, or the \textit{majority} of the team members not following proper software development practices or not being open to change. Moreover, how team members are \textit{dependent} on each other (\eg how much the outcome and speed of the code review affect their work) and getting stuck with people who are hard to cooperate with can also lead to conflicts. Misunderstandings among team members are also the main challenge during code review~\cite{bacchelli2013expectations}. Issues during communication might arise when team members' ability to explain and understand different perspectives is low. As one developer put it: \qud{It becomes complicated when there is, of course, a misunderstanding because there is one person who thinks about one reality and the other person has another reality, and the two don't match, and you maybe don't understand why the other person is making jokes and why the other person is screaming at you.}{17}

  Developers also report the importance of \textit{constructive feedback}, which has the following characteristics: (1) understandable (\ie explains reasons for decisions and suggestions with supporting examples); (2) actionable (\ie specific on what to do and how to do it); (3) focused on the topic (\ie specific on what is wrong prioritizing vital issues); and (4) includes positive points. In addition, to facilitate the communication of ideas, developers mention the importance of using \textit{synchronous} and \textit{in-person} communication where the exchange becomes fast and more elaborate.

  \item[\textbf{Organization.}]
  A software development environment creates a setting in which interpersonal interaction happens. Developers in our sample (see Table \ref{tab:descriptives}) reported important differences, especially between stable teams \textit{internal} to a company and the OSS \textit{community}. \textit{Internal} teams are paid for their job. They have more strictly set deadlines, higher developers' dependency on company hierarchy, and their motivations are affected by the need to keep the job and income.  Moreover, internal team members mostly work in close contact with each other (\eg in-person meetings, informal communication over coffee), which provides: (1) more context information; (2) more communicational cues and opportunities for clarification; (3) long-term development of inter-personal relationships.
  One developer said: \qud{Most of the cases the conflicts for me are with colleagues rather than volunteers. I think it's also because you have way more interactions with your colleagues rather than with the volunteers, at least in the projects I am involved with. We have many volunteers, but every one of them is only contributing very little}{8}.

  \textit{Community} members contribute to an OSS project voluntarily. In a community, interactions form organically, and connections are loose. Due to the voluntary contribution to the project, there is less pressure on community members who can leave problematic projects.

  \textit{Organization size} with a possible increase in the hierarchical or physical distance among developers also introduces complexity in communication. Besides, the \textit{quality of leadership} in an organization is related to problems during code review. Understanding superiors who support the teams' needs and goals facilitate smooth code reviews. Moreover, an organization's \textit{status}, such as the potential for growth or being understaffed, determines how the organization needs to become appealing to developers and how developers feel motivated or stressed.

  \textit{Maturity of the software development process} facilitates clarity of communication among developers. Moreover, automated processes, proper testing, following standards or conventions, and using tools (\eg static analyzers) can resolve some issues in advance, preventing the waste of resources during code reviews: \qud{If people discuss things that can be automated, it means their process is not set up efficiently \ldots~It's just my vision of how not to waste too much resources on things you don't really have to waste resources on}{14}.

  \textit{Organizational culture} provides for developers an agreement on how to communicate with and treat each other during code reviews. Besides valuing high-quality code, improvement, and innovation, developers mentioned the following values that organizations can incorporate into their culture: equality, respect, fairness, trust, tolerance, safety, openness, and cooperation rather than competitiveness. One interviewee stated: \qud{it's not forbidden to make mistakes in our team so if someone makes a mistake, he is not afraid that someone else will tell him <<hey, you made a mistake you are an idiot,>> but more like <<hey, I think we could do this in another way>>}{7}.

\end{description}

\subsection{\rqFour}
We describe the strategies developers employ to prevent and manage conflicts. We also present a detailed categorization of these conflict prevention and management strategies in Table \ref{tab:strategies}.

\begin{table}
\centering
  \caption{RQ5: Strategies to Prevent and Resolve Conflicts}
	\label{tab:strategies}
  \begin{adjustbox}{width=\textwidth}
    \begin{tabular}{lp{3.5cm}|lp{3.5cm}}

  \textbf{Strategies}                                     & \textbf{Participant ID}              							& \textbf{Strategies}                                                	& \textbf{Participant ID}   \\ \hline
  \multicolumn{2}{l|}{\textbf{Acknowledging the   conflict}}                                               & \multicolumn{2}{l}{\textbf{Adapting to individuals}}                                \\ \hline
  Addressing the conflict                              & D1, D6, D14, D15, D17, D18    						& \cellcolor[HTML]{EFEFEF}Individual approach                                              	&  \cellcolor[HTML]{EFEFEF}D4, D13, D15, D17, D18, D20           \\
  \cellcolor[HTML]{EFEFEF}Taking a break                                          & \cellcolor[HTML]{EFEFEF}D1, D6, D11, D13, D14, D17    							& Showing interest                                                  	&  D4, D13, D15      \\
  Opening the issue up for a discussion       &  D1                                           							& \cellcolor[HTML]{EFEFEF}Trying to understand the feelings of others         	&  \cellcolor[HTML]{EFEFEF}D15, D18           \\
  \cellcolor[HTML]{EFEFEF}Expressing needs and opinions                 &  \cellcolor[HTML]{EFEFEF}D8, D13-D15, D17, D18, D22  						& Adjusting own behaviour to others                      	&  D18              \\
  Expressing emotions                                 &  D4 ,D14, D17, D18, D22          						&                                                                            	&                                       \\
  \cellcolor[HTML]{EFEFEF}Reporting issues to authorities                  &  \cellcolor[HTML]{EFEFEF}D9, D17, D18                           						&                                                                            	&                                       \\ \hline
  \multicolumn{2}{l|}{\textbf{Active involvement   in conflict resolution}}       & \multicolumn{2}{l}{\textbf{Automation and standardization}}                         \\ \hline
  Being active towards   resolution               &  D17                                           						& \cellcolor[HTML]{EFEFEF}Create and update standards and conventions    		&  \cellcolor[HTML]{EFEFEF}D7, D11, D14, D16, D18-D20            \\
  \cellcolor[HTML]{EFEFEF}De-escalation                                           &  \cellcolor[HTML]{EFEFEF}D9, D17                                    						& Automate what is possible                                  		&  D7, D11, D13, D14, D16, D18, D19, D21 \\
  Formulating solutions                               &  D20                                          						& \cellcolor[HTML]{EFEFEF}Assign a responsible                                           	&  \cellcolor[HTML]{EFEFEF}D14                                   \\
  \cellcolor[HTML]{EFEFEF}Formulating lessons learned                     &  \cellcolor[HTML]{EFEFEF}D20-D22                                  						& Keep workflows flexible                                      		&  D14, D16                              \\ \hline
  \multicolumn{2}{l|}{\textbf{Preparation}}                                       & \multicolumn{2}{l}{\textbf{Third-party intervention}}                               \\ \hline
  Getting to know the code                         &  D1                                             						& \cellcolor[HTML]{EFEFEF}HR intervention                              					    	&  \cellcolor[HTML]{EFEFEF}D9, D18                               \\
  \cellcolor[HTML]{EFEFEF}Preparing what and how to communicate &  \cellcolor[HTML]{EFEFEF}D17                                           						& Giving a decision                                                		&  D17, D19                              \\
  Clarify differences                                    & D6, D7-D9, D15-D17                 						& \cellcolor[HTML]{EFEFEF}Managing work effort                                         		&  \cellcolor[HTML]{EFEFEF}D1, D8, D9, D11, D17                  \\
  \cellcolor[HTML]{EFEFEF}Do not develop ideas in isolation             &  \cellcolor[HTML]{EFEFEF}D9                                             						& Moderating communication                                		&  D8, D17, D18                          \\
  Plan steps                                                &  D1                                             						& \cellcolor[HTML]{EFEFEF}Setting priorities                                                 		&  \cellcolor[HTML]{EFEFEF}D10                                   \\
  \cellcolor[HTML]{EFEFEF}Agree on implementation                        &  \cellcolor[HTML]{EFEFEF}D7, D9                                       						& Team reorganization                                           	&  D11, D16, D20                         \\
  Send a change suggestion                       &  D9, D19                                     						& \cellcolor[HTML]{EFEFEF}Providing explanations                                        	&  \cellcolor[HTML]{EFEFEF}D6                                    \\
  \cellcolor[HTML]{EFEFEF}Clear requirements                                  &  \cellcolor[HTML]{EFEFEF}D1, D2, D7, D8                                                  &                                                                            	&                                       \\ \hline
  \multicolumn{2}{l|}{\textbf{Setting a process   against stagnation}}   & \multicolumn{2}{l}{\textbf{Training}}                                               \\ \hline
  Creating space for discussion                  &  D1, D2, D6, D9, D11, D16, D17, D19, D20        & \cellcolor[HTML]{EFEFEF}in Code of Conduct                          						&  \cellcolor[HTML]{EFEFEF}D5, D18, D21                          \\
  \cellcolor[HTML]{EFEFEF}Deciding on a team level                    	 	&  \cellcolor[HTML]{EFEFEF}D1, D11, D13, D16, D17, D21             				& in Company standards                        						&  D5, D18, D21                          \\
  Synchronizing                         					&  D6, D7, D9-D11, D15-D19, D21			            & \cellcolor[HTML]{EFEFEF}in the Software system                      						&  \cellcolor[HTML]{EFEFEF}D5                                    \\
  \cellcolor[HTML]{EFEFEF}Setting priorities                    					&  \cellcolor[HTML]{EFEFEF}D9, D11, D12, D16, D17, D19, D20        			& in Constructive feedback                    						&  D9, D17                               \\
  Gathering more knowledge              			&  D1, D4-D8, D11, D12, D14-D18, D20       		& \cellcolor[HTML]{EFEFEF}in Management                               							&  \cellcolor[HTML]{EFEFEF}D13                                   \\
  \cellcolor[HTML]{EFEFEF}Assigning new tasks                   				&  \cellcolor[HTML]{EFEFEF}D6                                      								&                                             									&                                       \\
  \end{tabular}
\end{adjustbox}
\end{table}

\begin{description}[leftmargin=0.3cm]

  \item[\textbf{Acknowledging the conflict.}]
  Developers mention the importance of acknowledging conflicts when they occur. Identifying conflict participants' emotions, needs, and goals helps clarify the situation and initiate discussions to find solutions. However, depending on the emotions' intensity, developers can postpone the discussions or next action steps. One developer said: \qud{The way we have to sort it is to take him for a date in the parlor downstairs and discuss it there in a more calm way like take out some stress, go down, have a beer and discuss it with him}{18}.

  \item[\textbf{Active involvement in conflict resolution.}]
  Developers initiate de-escalation procedures, acquire knowledge by asking questions, and formulate solutions to resolve conflicts constructively. Developers also formulate lessons learned by explicitly stating the potential benefits of the situation. \qud{This was our way to cope with negativity. So it was to interact more and ask for more details and ask for more information why is this so bad so that we have a learning moment from the community}{20}.

  \item[\textbf{Adapting to individuals.}]
  To prevent and address conflicts, developers try to understand the other developers' needs and feelings and adjust their behavior accordingly. The ability to adjust to individual needs is related to how close the developers are to each other and underpins the importance of good relationships within the team.
  One interviewee explained: \qud{There is a small number of people that you have to do reviews with. So that you can adapt somehow to their way of communication and you understand when they are getting angry, you understand how they work as people so how to calm them down, how to convince them, you understand which arguments work and which arguments they don't like.}{17}

  \item[\textbf{Automation and standardization.}]
  As mentioned in Section~\ref{sec:conflict_focus}, our findings also indicate a significant amount of conflicts on code readability (\ie code style, adherence to the code style conventions, consistency of the code style throughout the codebase). Moreover, some conflicts emerge about \textbf{bad procedures} (\eg not following standards, guidelines, and conventions on the overall software development process). Developers report that adopting standards and automation can help prevent and resolve conflicts. One developer said: \qud{The team should be composed in such a way that you automatize maximum of possible things. Starting with the styling and going as far as you can. And if this is not done properly, it can spin more and more conflicts.}{11}
   Even though automation and standardization facilitate communication, the workflows should not become rigid. Furthermore, developers also reported that it is helpful to have team members assigned to monitor the state and assess the need to update the standards, automation, and procedures used in the team.

  \item[\textbf{Preparation.}]
  To align the team's vision about the code change and prevent potential communication issues, developers prepare themselves for code review by clarifying the requirements and agreeing in advance on the code change implementation. Preparing for what and how to communicate during code reviews can also facilitate the prevention and management of conflicts during code reviews. An interviewee explained: \qud{I see the code review as some sort of the last step in the overall human interaction around developing the project. So many steps need to happen before in order to bring people on the same page.}{9}

  \item[\textbf{Set a process against stagnation.}]
  To avoid or resolve conflicts, developers set a process to resolve situations when the discussion is stagnating, for example, by taking discussion from code review tools into in-person meetings to align needs and expectations and give feedback on how the team members should cooperate. To avoid lengthy discussions over the review tools, developers also employ pair programming. Another strategy is to make decisions based on set priorities (\eg code simplicity over complexity, authors' opinion over alternatives on par, solution requiring the least effort). When the discussion stagnates, developers also gather knowledge from external resources, documentation, project history, and other developers to support their claims or get needed expertise. To obtain decision power, developers also involve the whole project team or field experts. Moreover, developers can employ democratic voting or list the pros and cons of proposed solutions to facilitate decision-making.

  \item[\textbf{Third-party intervention.}]
  In some cases, third parties other than conflict participants get involved in the conflicts' prevention and resolution. Developers sometimes involve Human Resources (HR) to resolve serious situations.
  Managers can also get involved and clear the situation by:
  (1) making the final decision, (2) providing rationale behind the decisions made, (3) setting priorities for the software development process (\eg code changes to implement/review), (4) moderating communication among conflict's participants, (5) adjusting workload among team members, (6) assigning responsibilities to individual team members, (7) providing instruction on how to proceed, and (8) reorganizing the team.

  \item[\textbf{Training.}]
  To get developers on the same page, organizations offer training on the software system, code of conduct, and conventions during onboarding. Developers also mention the usefulness of training on management and giving constructive feedback. \qud{I think the only way to actually sort this out is to explicitly train people to give constructive feedback all the time}{9}.
  
\end{description}

\section{Discussion}
\label{sec:discussion}
In this section, we discuss the main findings of our study and their implications for practitioners, researchers, and educators, as well as tool design. We first summarize our main findings:

\begin{description}[leftmargin=0.3cm]

    \item[\textbf{Conflicts are common and perceived as natural.}]
    From RQ1, we found that 21 developers (\ie all but one) experienced conflicts during code reviews. Nine stated to perceive conflicts as a natural and expected part of code review, and that code review may be the only environment for interpersonal conflicts at work.

    Developers reported several reasons for the conflict-prone nature of code reviews (Section~\ref{sec:rq0}) particularly that (1) developers address personally sensitive issues (\eg developers' skills, quality of developers' code) in an impersonal environment  (\eg code review tools, emails, task managers) and (2) discuss technical issues through a socially complex process. These two complementary angles suggest that conflict resolution strategies should focus on code reviews' technical and social nature. Similar difficulties to point (1) have been reported for distributed teams that need to communicate over distance and use tool-supported ways of communication~\cite{hinds2003out}.

    Even though these findings could be limited to our sample developers, we have no reason to think they are caused by sampling bias (we used unrelated sampling criteria--Section~\ref{sec:sample}). Therefore, it seems reasonable to derive that interpersonal conflicts during code reviews are an important topic to handle from both a research and a practical perspective.

    \item[\textbf{Conflicts are not to be prevented but managed.}]
    Previous studies~\cite{egelman2020pushback, schneider2016differentiating, squire2015floss} mainly focused on the negative aspects of code review conflicts. Our results align with those by \citet{barki2001interpersonal} in their investigation on conflicts in IS development. In particular, that conflicts are related to a decreased developers' productivity and well-being, lower process satisfaction and system quality, as well as worse adherence to budget, schedule, and requirements. However, our findings for RQ1 and RQ3 indicate that interpersonal conflicts during code review can also be constructive. In particular, our analysis revealed how conflicts could be a learning opportunity and bring novelty and additional value.

    Both positive and negative consequences of conflicts are closely related to the fulfillment of code review goals such as finding defects, code improvement, and knowledge transfer~\cite{bacchelli2013expectations}. On the one hand, all of the review goals are threatened by the presence of conflicts; on the other hand (as also reported in RQ1), conflicts might be a means to actually achieve them in an environment lacking an exchange of competing ideas.

    Developers actively use strategies to prevent and address conflicts when they are happening. These strategies show similarities with constructive conflict resolution strategies proposed in psychology and management literature (Section~\ref{sec:management}), such as active involvement in conflict resolution or setting priorities~\cite{coleman2012constructive}. However, some strategies, such as automation and standardization, are specific to software development and particularly to code review.

    \citet{huang2016effectiveness} emphasized the effectiveness of constructive suggestions as a conflict management strategy. Developers in our study also mention the importance of constructive feedback as a factor involved in the conflict dynamics. Our interviewees also propose ways of conflict management that lead to active involvement in situation resolution and clearing out priorities and next steps in the review and software development process.
    These findings, together with the evidence that code review conflicts happen naturally and developers cannot avoid them, indicates that further research can be designed and conducted on constructive conflict resolution specific to the context of code review.

    \item[\textbf{Code review and its context are strongly intertwined.}]
    When it comes to interpersonal conflicts, code review and its context (\ie code, developer, team, organization) are strongly intertwined. Indeed, as reported when answering RQ3 (Section~\ref{sec:RQ3}), social interaction problems during code review can negatively affect code, developers, development teams, and organizations in addition to the code review process. These findings align with Hartwick and Barki~\cite{barki2001interpersonal} by showing that conflicts result in decreased developers' productivity and well-being, lower process satisfaction, system quality, worse adherence to budget, schedule, and requirements.

    Furthermore, as a response to RQ4 (Section~\ref{sec:factors}), we reported factors related to code, developer, team, and organization that can be fruitful areas of intervention to support social interactions. Our findings describe many technical and non-technical factors identified in literature that play a role in software development, influence source code quality or code change acceptance such as: (1) code and review related factors (\eg patch size, patch priority, code complexity~\cite{baysal2013influence}, understandability~\cite{scalabrino2017automatically}, requirements quality~\cite{liu2011relationships}, documentation clarity~\cite{ciurumelea2020suggesting}); (2) developer-related factors (\eg experience, engagement~\cite{baysal2013influence}) and (3) team and organisation level factors (\eg relationship quality~\cite{tsay2014influence}, leadership~\cite{wickramasinghe2015diversity}).

    Conceptually, destructive interpersonal conflicts are a form of community smells, which are communication issues leading to communities' inability to tackle software development problems. Research has shown a relationship between community smells and code smells~\cite{palomba2018beyond}, similarly to the consequences of conflicts for the codebase as well as the team and organization that we found (Section~\ref{sec:RQ3}).
    Conflicts can also lead to sub-optimal decisions and solutions regarding performance, security, or maintainability resulting in technical debt~\cite{rios2018tertiary}. Moreover, team and organization-level factors listed in Section~\ref{sec:factors} can contribute to social debt, which is the additional cost that occurs when strained social and organizational interactions get in the way of smooth software development and operation~\cite{tamburri2015social}. Therefore, our study corroborates that further research on the role of the reported factors can bring new insights into the relation between community smells and code smells and between social and technical debt.

    \item[\textbf{Conflicts in code review are also domain-specific.}]
    Our results for RQ1 indicated that conflicts during code reviews are \textit{non-technical conflicts about technical issues} and that code review generates also conflicts that are specific only to this context. For this reason, even though some conflict management strategies can be borrowed from psychology and management literature (\eg acknowledging conflicts and adapting to individuals), as reported in RQ5, additional strategies---specific to code review conflicts---must be investigated and brought into practice (such as automation and standardization). In other words, existing off-the-shelf solutions/approaches to handle conflicts are not enough for the code review context.

\end{description}

\subsection{Decomposing Conflicts}
To understand and navigate conflicts successfully, it is crucial to decompose them into their individual components. This decomposition enables the understanding of the situation on a more fine-grained level and the identification of strategies to address specific and situational needs for conflict prevention, management and resolution.
One of the main goals of our study (answered in RQ2) was to describe conflicts during code review using the two-dimensional construct by \citet{hartwick2002conceptualizing}, which consists of \textit{conflict focus} and \textit{conflict components} (\ie \textit{disagreements}, \textit{interference}, \textit{emotions}). Unlike disagreement and emotions, interference (\ie behaviors, symptoms) is observable, directly serving as indicators of more risky situations that can lead to conflicts. However, the behaviors of others are subject to the interpretation of an individual. If the individual has no negative emotions, there are no conflicts~\cite{hartwick2002conceptualizing}.  Therefore, it is crucial to look for early signs of emotions for conflict resolution and management besides interference.  Understanding emotions can help devise solutions to prevent their occurrence, hence conflicts in the future. To summarize, our findings can help:

\begin{enumerate}
\item identify areas needing improvement or additional knowledge for the review and communication to be successful through analyzing \textit{focus of conflict};
\item understand the \textit{disagreement} where developers' cognition differs to unify or clarify their perspectives for effective conflict resolution;
\item analyze developers' communication during code reviews to detect signs of communication difficulties or inappropriate behaviors (\ie \textit{interference}) to  detect the risk of conflicts happening;
\item conduct sentiment analysis~\cite{feldman2013techniques} on data gathered from code review tools to detect and identify negative \textit{emotions}, which can provide clues on the developers' needs that can be addressed for conflict resolution or prevention.
\end{enumerate}

\subsection{Implications}
The results presented in this paper can be translated into implications for practitioners and management, as well as for educators and researchers. The results can also help inform the design of tools that better support the communication among developers during code reviews.

\subsubsection{Practitioners}
\label{sec:practitioners}
Our findings present (1) various forms of conflicts (\ie communication issues, symptoms, interfering behavior, differences in perspectives), (2) factors related to conflicts, and (3) strategies for conflict prevention and management that can serve practitioners as an inspiration on how to detect and prevent or manage conflicts during code review.
We also identified that developers' technical skills (\eg program comprehension and analysis) and soft skills (\eg communication, cooperation, engagement in constructive criticism) described in the literature~\cite{linck2013competence} could facilitate the prevention, management, and resolution of conflicts. Therefore, training to improve these skills has the potential to prevent conflicts.
Organizations should offer practitioners training during onboarding on the software system, conventions, and standards used in the organization for software development to minimize conflicts. Moreover, organizations should design and conduct training that improves practitioners' soft skills (\eg communication, cooperation, giving constructive feedback).

\subsubsection{Tool Design}
Tools can help identify and mitigate conflicts during code review.

\begin{enumerate}
\item \textit{Provide conflict risk metrics.} This study identified numerous symptoms of conflicts and related factors that can be used to build a metric suite to facilitate the detection of conflict-prone code review environments. Some of the symptoms we identified relate to the code and change (code and documentation quality, review complexity), and some relate to the characteristics of the communication (lack of progress, constructive feedback), team (size, composition), project, or their review process (frequency of reviews). Analyzing these factors can bring insights into how conflict-prone a specific review is. \citet{egelman2020pushback} use such metrics to identify pushback in code reviews. Furthermore, text analysis (\eg Sentiment Analysis~\cite{feldman2013techniques}) can be used to estimate the emotions in the review and provide recommendations on which reviews or teams need communication improvement, conflict resolution, or management intervention. Furthermore, the text-based computational analysis offers mechanisms to study conflicts from other angles, using the content of the text and employing different machine learning techniques~\cite{maerz2020text}.
\item \textit{Detect stagnation points.} Conflicts can be indirectly observed through discussion lacking progress. Tools can act as a facilitator of progress by detecting stagnation in the discussion and nudging developers in using some strategy (see Table \ref{tab:strategies}) to resolve the stagnation point.
\item \textit{Support switching to synchronous communication.} Code review conflicts are dependent on mutual understanding and can get personal, but they happen through tools and asynchronous communication. Developers mention the importance of discussing more complicated issues in person and having faster feedback on other developers' understanding. Therefore, tools can acknowledge this shortcoming by directly requesting a meeting or a synchronous peer review.
\item \textit{Assist in expressing priorities and issue tracking.} Conflicts in code review arise based on misalignment in goals and priorities related to the change and task. Tools can further support the expression of importance, goals, and priorities of the review and issues detected in it, for example, by assigning priorities to reviews and comments.
\end{enumerate}

\subsubsection{Educators}
Code reviews should be an integral part of software project courses. While designing project courses, educators should also consider the points we summarized for practitioners above (\eg preparations, simplicity and clarity, cooperation). For instance, conducting code reviews during course projects can serve as an experiential learning technique~\cite{patrick:2011} to facilitate students' communication and collaboration skills besides technical skills (\eg requirements elicitation, software design, coding). Furthermore, students need support to develop sufficient communication skills to formulate and defend technical decisions and provide feedback to their colleagues in a constructive way. Software project courses should also provide students with the practice of adhering to software development guidelines, standards, and coding conventions.

Developers reported the importance of using social skills (\eg communication, cooperation) as factors involved in interpersonal conflicts (see Section ~\ref{sec:factors}). Developers find \textbf{training} on giving constructive feedback and other topics useful for conflict prevention, resolution and management. Literature suggests that developers need the ability and willingness to share knowledge, constructively criticize, make and fulfill agreements in the team and learn from other team members ~\cite{linck2013competence}. Skills can be trained to improve productivity and collaboration including conflicts~\cite{coleman2012constructive}.

\subsubsection{Researchers}
Conflicts' consequences can severely affect not only the code review process but also a codebase, any involved developers and teams, as well as an entire organization  (Section~\ref{sec:factors}).
Based on these findings, an interesting research path is to investigate interpersonal conflicts during code review through the lens of the \textit{gestalt field theory} in psychology literature~\cite{parlett1991reflections}.  Field theory is the observation that any situation's meaning is determined by the relationship between the thing we are focusing on (\eg code review) and the context it occurs within (\eg code, developer, team, organization). According to this theory, it is not sufficient to fix individual factors to correct a dysfunctional situation. Rather the solution lies in finding a new, functional equilibrium in the field. Therefore, fixing interpersonal conflicts would require finding a balance between the code, review, developer, team, and organization-specific factors that facilitate a constructive discussion in a stable environment.

\subsection{Future Work}
This study acted as an initial broad description of potential forms of conflicts during code review, their context, and effects. There are several areas that require further research work to understand conflicts and mitigate their potentially negative consequences:

\begin{enumerate}
	\item Quantitatively confirm the conflicts' frequency and severity in code reviews.
	\item Quantitatively investigate positive and negative consequences of conflicts with a focus on their relation to the fulfillment of the review goals.
	\item Further describe the mechanisms of involvement of individual factors.
	\item Investigate the effectiveness of in-person modes of communication and their integration in code review tools.
	\item Investigate whether developers are sufficiently informed and equipped to handle conflicts during code reviews.
	\item Use proposed components and factors to build conflicts risk assessment metrics.
	\item Quantitatively investigate the efficiency and effectiveness of conflict management strategies developers use and their applicability.
	\item Identify further conflict management strategies in the literature and assess their utility in the context of code review.
\end{enumerate}

\section{Conclusion}
In this study, we presented a qualitative investigation on interpersonal conflicts in code review. By interviewing 22 developers with a range of expertise and from different environments, we found that conflicts are common in code review and perceived as a natural part of the task. However, not all conflicts lead to poor outcomes--some improve various social and technical aspects of software development. Therefore, conflicts do not necessarily need to be prevented but rather managed. Developers describe a number of factors that affect interpersonal conflicts during code review, as well as several problem-focused strategies that can be used to identify and understand conflicts, thus leading to better conflict management. We hope that the insights we have discovered and presented in this paper will form the basis for further research on this prominent aspect of code review and software engineering.

\begin{acks}
P. Wurzel Gon\c calves and A. Bacchelli gratefully acknowledge the support of the Swiss National Science Foundation through the SNF Project No. PP00P2\_170529.
\end{acks}

\bibliographystyle{ACM-Reference-Format}
\bibliography{cr_conflicts}

\appendix

\section{Interview structure}
\label{sec:appendixI}
We conducted 22 semi-structured interviews. The duration of the interviews varied from 45 to 60 minutes. Each interview followed the structure presented below. The interviewer followed up on the respondents' narratives in the following cases: (1) to contribute to the extension of knowledge about conflicts; and (2) to understand the topics present in previous interviews or uncover new areas to explore.
We paid particular attention to areas where the respondents' experience and opinions differed from other respondents.\smallskip

\textbf{Introduction} (5 min)
\begin{itemize}
\item Welcome
\item Information about the research and the interview
\item Reminder of rights and consent with participation
\item Asking for the context of work with code review
\end{itemize}
\smallskip

\textbf{Body} (35 \textendash 50 min)
\begin{itemize}
\item Experience with conflicts and examples
\item How do the conflicts look like?
\item What are the conflicts about?
\item How do the conflicts influence them, their work, and their team?
\item Are there any positive or negative consequences of conflicts?
\item Why do the conflicts happen?
\item What are the things that influence whether and how the conflict happens?
\item What would help them prevent and manage conflicts?
\end{itemize}
\smallskip

\textbf{Ending} (5 \textendash 10 min)
\begin{itemize}
\item Space for questions
\item Final remarks and summary
\item Offering results and publications
\end{itemize}

\section{Qualitative analysis specification}
\label{sec:appendixV}

\subsection{Bottom-up approach}
The bottom-up approach is an inductive qualitative analysis approach where the researcher derives codes and themes from the data content, closely matching the data's content. It differs from the top-down (deductive) method in which the researcher brings in advance a series of concepts, ideas, or topics to the analysis. However, even if the analysis is purely bottom-up, researchers may bring into the analysis at least a set of concepts and ideas that help facilitate coding~\cite{clarke2015thematic}.

Our analysis specified the topics and areas of interest based on literature and concepts related to interpersonal conflicts. The developers' narratives drove the codes themselves and the themes formed from the codes as well as the terms they have brought up in the interviews.

We chose the ``bottom-up approach" during the Thematic Analysis for the following reasons: (1) Conflicts in code reviews are an emerging and poorly understood topic; (2) We conduct an exploratory study of conflicts as applied in code review specifically, thus we want to keep the analysis as close to the experience of developers as possible; (3) We aim to provide results that are beneficial to the developers, and deriving themes based on developers' narratives keeps the coding relevant to the audience; (4) We aim to mitigate the researcher's bias by preventing the researcher from bringing concepts and ideas that do not necessarily link to the semantic data content.

\subsection{Peer-debriefing}
Peer debriefing is a review of the research process conducted by a researcher external to the analysis yet familiar with the data or with the phenomena under investigation. The reviewer supports the analyst, challenges the researchers’ assumptions, and provides critical feedback on the analysis’s progress~\cite{creswell2000determining}.

During the initial coding phase of the thematic analysis (Step 2 in Section \ref{sec:met:analysis}), to discuss the coding scheme, the first author had meetings regularly (weekly or bi-weekly) with the second author, who had familiarized with the data by reading the complete set of interview transcripts.
These meetings aimed to introduce the second author to the current structure of codes. The analyst (the first author) could ask clarifying questions about possible meaning and interpretation of data or software engineering practices and terminology. The reviewer (the second author) had the opportunity to ask for codes' explanation, to point out inconsistencies in the analysis, to provide observations about the data, and to relate to the current state and potential extensions of the analysis.
All three authors discussed the outcome of the peer debriefings in a meeting before the third author proceeded with the audit.

\subsection{Audit-like process}
The audit trail is a procedure to validate the qualitative analysis results. It requires the researcher to provide detailed information on how the analysis was conducted to auditors external to the analysis. A formal audit aims to examine both the inquiry's process and product and determine the findings' trustworthiness.
Typically, an external auditor, such as a thesis reviewer, conducts the audit. However, the auditor in this study was the third author. Therefore, the audit's position towards the analysis is not impartial due to the auditor's familiarity with the project.  Furthermore, the third author took part in the study's methodological and writing process and discussions. Therefore, we call the audit in this study an ``audit-like'' process---an audit performed by a person external to the analysis yet internal to the project. An audit-like process is also among the methods used to enhance the trustworthiness of qualitative studies~\cite{carcary2009research}.

The audit was performed on several documents, including a text summarizing the related work, methodology, analysis results, the analyst's notes, theme definitions, the NVivo file with complete coding, and interview transcripts.
The auditor read three interviews with abundant coding to familiarize themselves with at least a third of the relevant data. Then the auditor proceeded with the documents covering the analysis, results, coding, notes, and definitions of themes.

The audit aimed to reach the following goals:
\begin{itemize}
\item to get acquainted with the methodology and results of the analysis and related documentation
\item to review whether the analysis results are a good representation of the data
\item to review whether the analysis does not breach knowledge available to software engineering unless well supported
\item to review whether the results are beneficial for the community
\item to control for issues and shortcomings of the analysis
\end{itemize}

The auditor proposed changes and improvements to the themes'  organization and reporting of the themes. However, no shortcomings in the validity of the analysis were identified in the audit.

\end{document}